\documentclass{ieeeaccess}

\usepackage[hyphens]{url}
\usepackage[hidelinks]{hyperref}
\hypersetup{breaklinks=true}
\usepackage{cite}
\usepackage{amsmath,amssymb,amsfonts}
\usepackage{algorithm}
\usepackage{algpseudocode}
\usepackage{graphicx}
\usepackage{textcomp}
\usepackage{afterpage}
\usepackage{array}
\usepackage{supertabular}
\usepackage{multirow}
\usepackage{multicol}
\usepackage{multirow}
\usepackage{afterpage}
\usepackage{siunitx}
\usepackage{color}
\usepackage{subfigure}
\usepackage{makecell}
\usepackage{txfonts}
\usepackage{silence}
\usepackage{gensymb}
\usepackage{bm}
\makeatletter
\newcommand{\thickhline}{%
    \noalign {\ifnum 0=`}\fi \hrule height 1pt
    \futurelet \reserved@a \@xhline
}
\newcolumntype{"}{@{\hskip\tabcolsep\vrule width 1pt\hskip\tabcolsep}}
\makeatother
\def\BibTeX{{\rm B\kern-.05em{\sc i\kern-.025em b}\kern-.08em
    T\kern-.1667em\lower.7ex\hbox{E}\kern-.125emX}}
\begin{document}
\history{Date of publication xxxx 00, 0000, date of current version xxxx 00, 0000.}
\doi{10.1109/ACCESS.2024.DOI}

\title{Enhancing eLoran Timing Accuracy via Machine Learning with Meteorological and Terrain Data}
\author{\uppercase{Taewon Kang}\authorrefmark{1},
\uppercase{Seunghyeon Park}\authorrefmark{1},
\uppercase{Pyo-Woong Son}\authorrefmark{2},
\uppercase{and Jiwon Seo}\authorrefmark{1, 3}, \IEEEmembership{Member, IEEE}}

\address[1]{School of Integrated Technology, Yonsei University, Incheon 21983, Republic of Korea}
\address[2]{Department of Electronics Engineering, Chungbuk National University, Cheongju 28644, Republic of Korea}
\address[3]{Department of Convergence IT Engineering, Pohang University of Science and Technology, Pohang 37673, Republic of Korea}

\tfootnote{
This work was supported in part by Grant RS-2024-00407003 from the ``Development of Advanced Technology for Terrestrial Radionavigation System'' project, funded by the Ministry of Oceans and Fisheries, Republic of Korea; 
in part by the Korean government (Korea Aerospace Administration, KASA), under Grant RS-2022-NR067078; 
in part by the National Research Foundation of Korea (NRF), funded by the Korean government (Ministry of Science and ICT, MSIT), under Grant RS-2024-00358298;
in part by the Unmanned Vehicles Core Technology Research and Development Program through the NRF and the Unmanned Vehicle Advanced Research Center (UVARC), funded by the MSIT, Republic of Korea, under Grant 2020M3C1C1A01086407;
and in part by the MSIT, Korea, under the Information Technology Research Center (ITRC) support program, supervised by the Institute of Information \& Communications Technology Planning \& Evaluation (IITP), under Grant IITP-2024-RS-2024-00437494.
}

\markboth
{Kang et al.: Enhancing eLoran Timing Accuracy via Machine Learning with Meteorological and Terrain Data}
{Kang et al.: Enhancing eLoran Timing Accuracy via Machine Learning with Meteorological and Terrain Data}

\corresp{Corresponding authors: Pyo-Woong Son (pwson@cbnu.ac.kr) and Jiwon Seo (jiwon.seo@yonsei.ac.kr)}

\begin{abstract}
The vulnerabilities of global navigation satellite systems (GNSS) to signal interference have increased the demand for complementary positioning, navigation, and timing (PNT) systems.
To address this, South Korea has decided to deploy an enhanced long-range navigation (eLoran) system---a radionavigation system that relies on high-power, low-frequency terrestrial signals---as a complementary PNT solution.
Similar to GNSS, eLoran provides highly accurate timing information, which is essential for applications such as telecommunications, financial systems, and power distribution.
However, the primary sources of error for GNSS and eLoran differ.
For eLoran, the main source of error is signal propagation delay over land, known as the additional secondary factor (ASF).
This delay, influenced by ground conductivity and weather conditions along the signal path, is challenging to predict and mitigate.
In this paper, we measure the time difference (TD) between GPS and eLoran using a time interval counter and analyze the correlations between eLoran/GPS TD and eleven meteorological factors.
Accurate estimation of eLoran/GPS TD could enable eLoran to achieve timing accuracy comparable to that of GPS.
We propose two estimation models for eLoran/GPS TD and compare their performance with existing TD estimation methods.
The proposed WLR--AGRNN model captures the linear relationships between meteorological factors and eLoran/GPS TD using weighted linear regression (WLR) and models nonlinear relationships between outputs from expert networks through an anisotropic general regression neural network (AGRNN).
The model incorporates terrain elevation to appropriately weight meteorological data, as elevation influences signal propagation delay.
Experimental results based on four months of data demonstrate that the WLR--AGRNN model outperforms other models, highlighting its effectiveness in improving eLoran/GPS TD estimation accuracy.
\end{abstract}


\titlepgskip=-15pt

\maketitle

\section{Introduction}
\label{sec:Intro}

The Global Navigation Satellite Systems (GNSS) are considered the most widely used positioning, navigation, and timing (PNT) systems today, owing to their accurate positioning and timing capabilities, as well as their convenience. 
However, due to the low received signal power, GNSS is vulnerable to high-power radio frequency interference (RFI) \cite{Roberts2022:Detection, Park21:Single, Moon24:HELPS, Kriezis2024:GNSS, Kim23:Low, Lee22:Evaluation}. 
Additionally, ionospheric anomalies can disrupt GNSS signals \cite{Pullen2009:Impact, Seo11:Availability, Seo14:Future, Lee17:Monitoring, Sun21:Markov, Lee2011:Observations, Lee22:Optimal}.

After repeated intentional high-power GPS jamming incidents from North Korea since 2010, South Korea decided to deploy an enhanced long-range navigation (eLoran) system as a complementary PNT solution to GNSS \cite{Son20:eLoran, Rhee21:Enhanced, Son23:Demonstration, Son24:eLoran}. 
eLoran, an enhanced and backward-compatible version of Loran, is a high-power, low-frequency groundwave-based radionavigation system \cite{Pelgrum06, Williams13}. 
Although its positioning accuracy is lower than that of GNSS, the high signal power of eLoran makes it robust to RFI.

A significant source of error for eLoran positioning and timing is the additional secondary factor (ASF), which refers to the propagation delay caused by the terrain along the groundwave propagation path \cite{Lo09, Zhou2013:A}.
The ASF consists of spatial and temporal components, referred to as spatial ASF and temporal ASF, respectively. 
The spatial ASF can be mitigated using ASF maps, and accurate methods for generating these maps have been actively studied \cite{Luo2006:ASF, Son19:Universal, Kim22:First, Williams2000:Mapping, Park20:Effect}. 
The temporal ASF can be corrected using temporal corrections generated by a nearby differential eLoran reference station, also known as a DLoran station \cite{Hargreaves2012:ASF, Van2014:eDLoran, Chang2023:Evaluation}. 
A DLoran station generates differential corrections and transmits the correction information to an eLoran transmitter via a data communication channel. 
The transmitter then modulates this correction information onto Loran pulses and broadcasts the modulated pulses to users. 
This data channel using Loran pulses is referred to as the Loran Data Channel (LDC) \cite{Li2012:Research, Yang2020:Experimetal, Lyu2021:Application, Son2023:Demonstration}.

An eLoran receiver can generate precise time information as a byproduct of its positioning process when signals from more than three transmitters are received. 
eLoran provides two-dimensional positioning, and three transmitters are sufficient to obtain both horizontal position and time information. 
If the eLoran receiver is at a known location, a single transmitter is sufficient for generating coordinated universal time (UTC)-synchronized pulse-per-second (PPS) signals. 
A conventional eLoran receiver is typically integrated with a GNSS receiver, allowing it to output PPS signals based on GNSS when available, and based on eLoran when GNSS is unavailable. 
Consequently, the time difference (TD) between eLoran- and GNSS-based PPS signals is nearly zero when the receiver switches its mode from GNSS to eLoran. 
In this case, the eLoran receiver does not need to be at a known location. 
Regardless of the receiver’s location, TD remains close to zero immediately after the mode switch from GNSS to eLoran.

\begin{figure}
  \centering
  \includegraphics[width=1.0\linewidth]{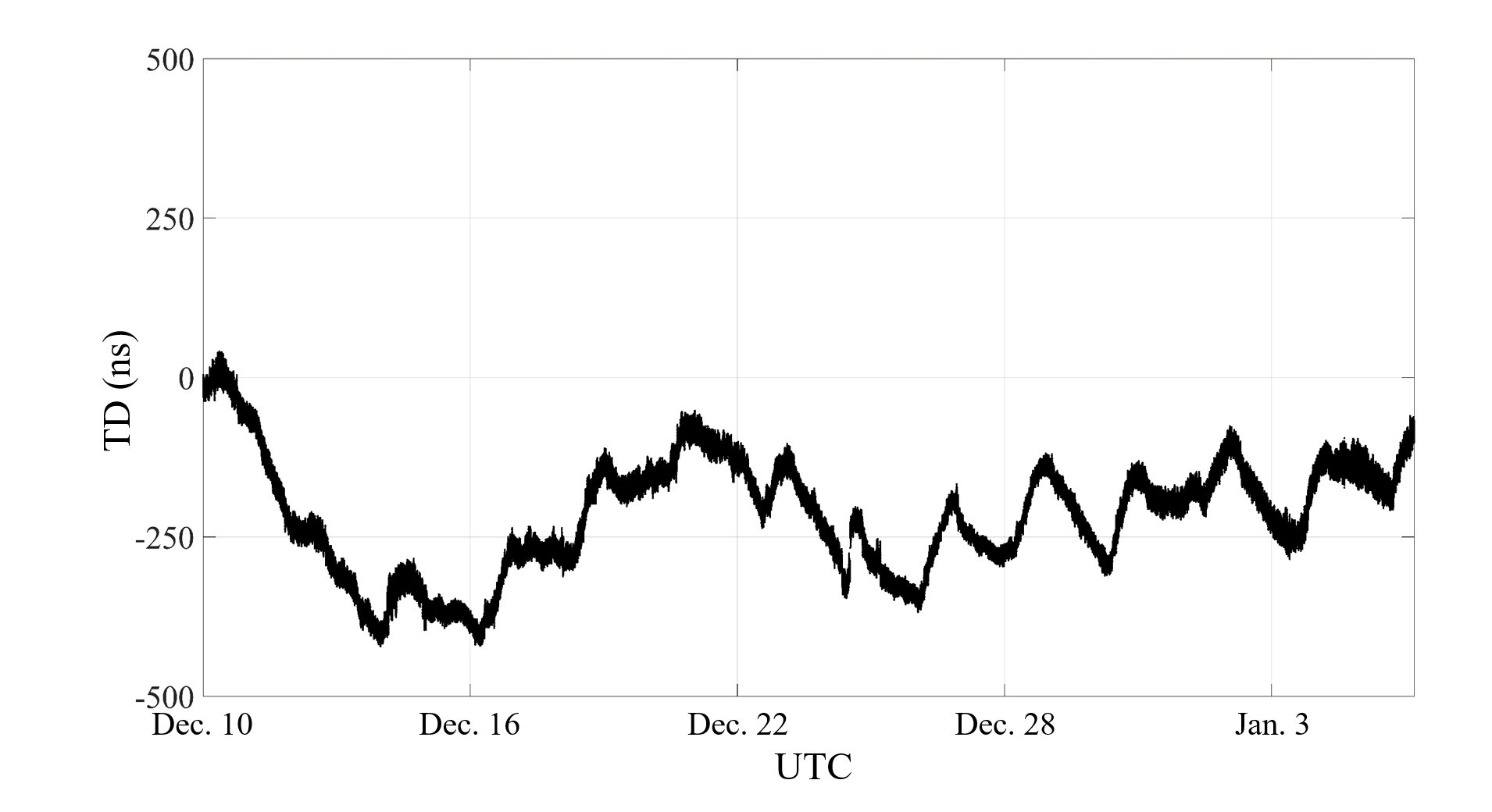}
  \caption{Measured eLoran/GPS time difference (TD) at a fixed location over time.}
  \label{fig:TimeDifference}
\end{figure}

However, as time elapses, the TD changes due to temporal variations of the ASF, as depicted in Figure~\ref{fig:TimeDifference}. 
To ensure reliable time information from eLoran PPS signals after switching from GNSS to eLoran---which is the focus of this study---it is crucial to accurately estimate and compensate for the temporal variation in TD. 
This allows TD to remain close to zero (i.e., the accuracy of eLoran timing closely approximates that of GNSS timing), even during periods of GNSS signal unavailability caused by RFI.

If a DLoran station is in proximity to a user, the user can leverage the provided differential correction information to mitigate the temporal ASF error and, consequently, maintain TD close to zero. 
However, DLoran stations are typically deployed at harbors to support maritime eLoran users, and deploying a large number of DLoran stations in inland areas to support eLoran timing users may be impractical. 
Furthermore, not all eLoran transmitters are equipped with LDC. 
For example, among the three transmitters in Korea, only the Socheongdo transmitter has LDC. 
Therefore, users who cannot receive signals from the Socheongdo transmitter are unable to utilize the differential correction information provided through LDC.

In an effort to mitigate temporal ASF error in cases where a DLoran station or a signal from a transmitter with LDC is unavailable, studies have been conducted to predict the temporal variations of propagation delay using meteorological data \cite{Lebekwe2021:Meteorological, Meng2009:ASF, Pu2024:Differential}, as it is well known that signal propagation delay is influenced by meteorological conditions along the propagation path. 
In particular, the use of neural network models that incorporate various meteorological inputs to estimate propagation delay has received increasing attention \cite{Pu2019:Analysis, Pu2021:Accuracy, Liu2023:eLoran}. 
However, these studies have shown limited estimation accuracy due to the use of only a small number of meteorological data sources from a few locations along the propagation path, and their estimation methods have potential for further improvement.

Moreover, these studies considered only meteorological data for model training, which may not adequately capture the full range of factors influencing the propagation of low-frequency (LF) groundwave signals such as eLoran. 
In particular, several studies investigating LF groundwave propagation over irregular terrain \cite{Zhou2011:LF, Wang2019:LF, Zhou2017:LF, Zhang2024:Time} have confirmed that topographic features such as terrain elevation---which cannot be inferred from meteorological data alone---have a significant impact on propagation delay. 
Based on these findings, we incorporate terrain elevation alongside meteorological data into TD prediction to further improve the timing accuracy of eLoran.

In this paper, we propose using seven meteorological factors and terrain elevation data to estimate the eLoran/GPS TD. 
Meteorological data along the groundwave propagation path are obtained by constructing meteorological data maps from measurements collected at a limited number of locations. 
We propose two estimation models that have not previously been applied to eLoran/GPS TD estimation: the least absolute shrinkage and selection operator (LASSO) regularized multivariate polynomial regression (MPR) model and the weighted linear regression--anisotropic general regression neural network (WLR--AGRNN) model. 
In particular, the WLR--AGRNN model is designed to effectively incorporate terrain elevation data alongside meteorological factors. 
The performance of the proposed models is validated through field experiments.

The contributions of this study are summarized as follows:
\begin{itemize}
\item Analysis of correlations:
The correlations between eLoran/GPS TD and meteorological factors are analyzed. 
Meteorological factors showing strong correlation with TD are incorporated into the proposed TD estimation models.

\item Meteorological data mapping:
Grid maps of meteorological data are constructed using measurements from a limited number of locations along the groundwave propagation path. 
This approach improves data accuracy at locations where direct measurements are not available.

\item Proposed estimation models:
Two learning-based eLoran/GPS TD estimation models are proposed. 
One of these models is specifically designed to account for the impact of terrain elevation along the propagation path.

\item Experimental validation:
The performance of the proposed models is validated using real-world measurements collected over a four-month period. 
This significantly exceeds the validation periods of approximately one week reported in previous studies \cite{Pu2019:Analysis, Liu2023:eLoran, Pu2021:Accuracy}.
\end{itemize}

This study enables more accurate eLoran/GPS TD estimation compared to previous research, addressing one of the key limitations of eLoran signals---their lower timing accuracy relative to GNSS. 
Several applications already utilize eLoran’s timing information as a backup to GPS. 
In the UK, multiple government projects that use eLoran to provide accurate timing information are currently underway \cite{Rivkin2024:Navigation}. 
In the US, the Department of Transportation has identified eLoran as a promising candidate for delivering resilient timing services as part of a broader GPS backup strategy, as outlined in its 2021 report on complementary positioning, navigation, and timing (PNT) technologies \cite{USDOT2021:ComplementaryPNT}. 
Additionally, the New York Stock Exchange (NYSE) hosted a demonstration of eLoran’s capability to deliver accurate timing within its offices \cite{Offermans2017:Providing}. 
These studies and applications demonstrate the practical value of enhancing eLoran timing performance in real-world scenarios where reliance on GNSS alone may be unreliable or infeasible. 
While our study validates the proposed network using data collected along a single propagation path, extending the approach to multiple paths with diverse training datasets could further enhance eLoran signal coverage and maximize its broader applicability.

The remainder of this paper is organized as follows. 
Section~\ref{sec:DataCollection} analyzes the correlations between eLoran/GPS TD and multiple meteorological factors and explains the data preparation process for TD estimation. 
Section~\ref{sec:EstModels} describes the proposed eLoran/GPS TD estimation models. 
Section~\ref{sec:Results} presents the estimation results of the proposed models along with comparisons to existing TD estimation methods. 
Finally, Section~\ref{sec:Conclusion} concludes the study.

\section{Proposed Set of Data for eLoran/GPS TD Estimation}
\label{sec:DataCollection}

In this section, we describe the data collection setup for measuring eLoran/GPS TD, along with the meteorological and terrain data used in this study. 
We then analyze the relationship between the recorded TD values and each of the eleven meteorological factors at the user receiver location. 
Based on this analysis, seven meteorological factors are selected for eLoran/GPS TD estimation. 
To obtain meteorological data along the groundwave propagation path, meteorological data maps are constructed using measurements from ten weather stations. 
Additionally, terrain elevation data along the propagation path is incorporated to improve estimation accuracy.

\subsection{Data Collection}

\begin{figure}
  \centering
  \includegraphics[width=0.9\linewidth]{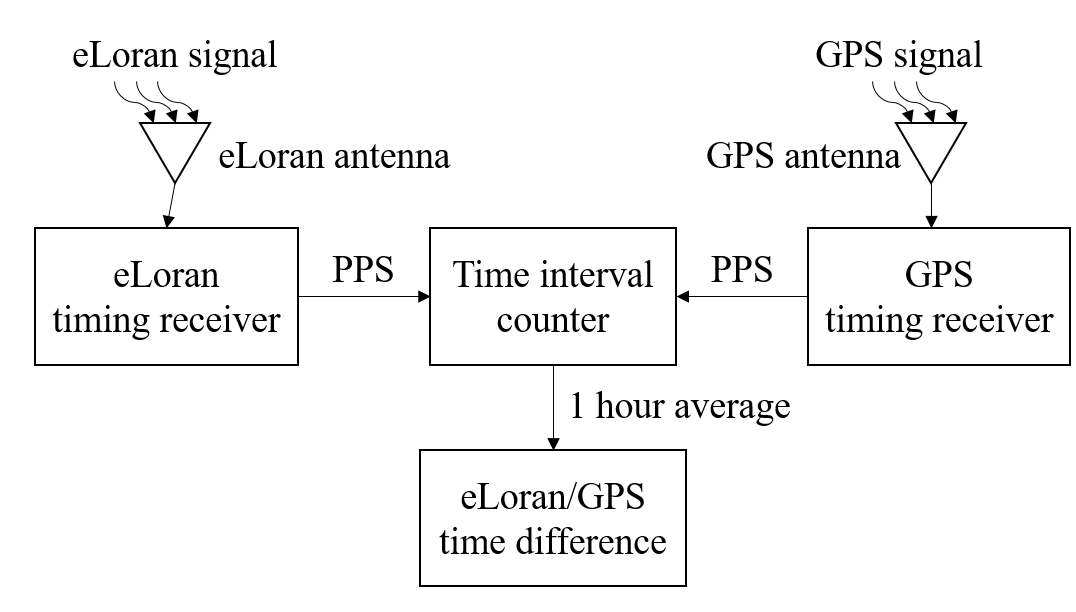}
  \caption{Block diagram of eLoran/GPS TD measurement system.}
  \label{fig:MeasureSys}
\end{figure}

Figure~\ref{fig:MeasureSys} shows the block diagram of the data collection setup used to measure eLoran/GPS TD values. 
The time difference between the PPS signals from the eLoran and GPS timing receivers was measured in nanoseconds using a time interval counter (TIC). 
The eLoran/GPS TD was recorded every second and stored on a computer along with other timing information. 
Since the available meteorological data have a one-hour resolution, the eLoran/GPS TD was averaged over one-hour intervals to match this resolution.

\begin{figure}
  \centering
  \includegraphics[width=0.9\linewidth]{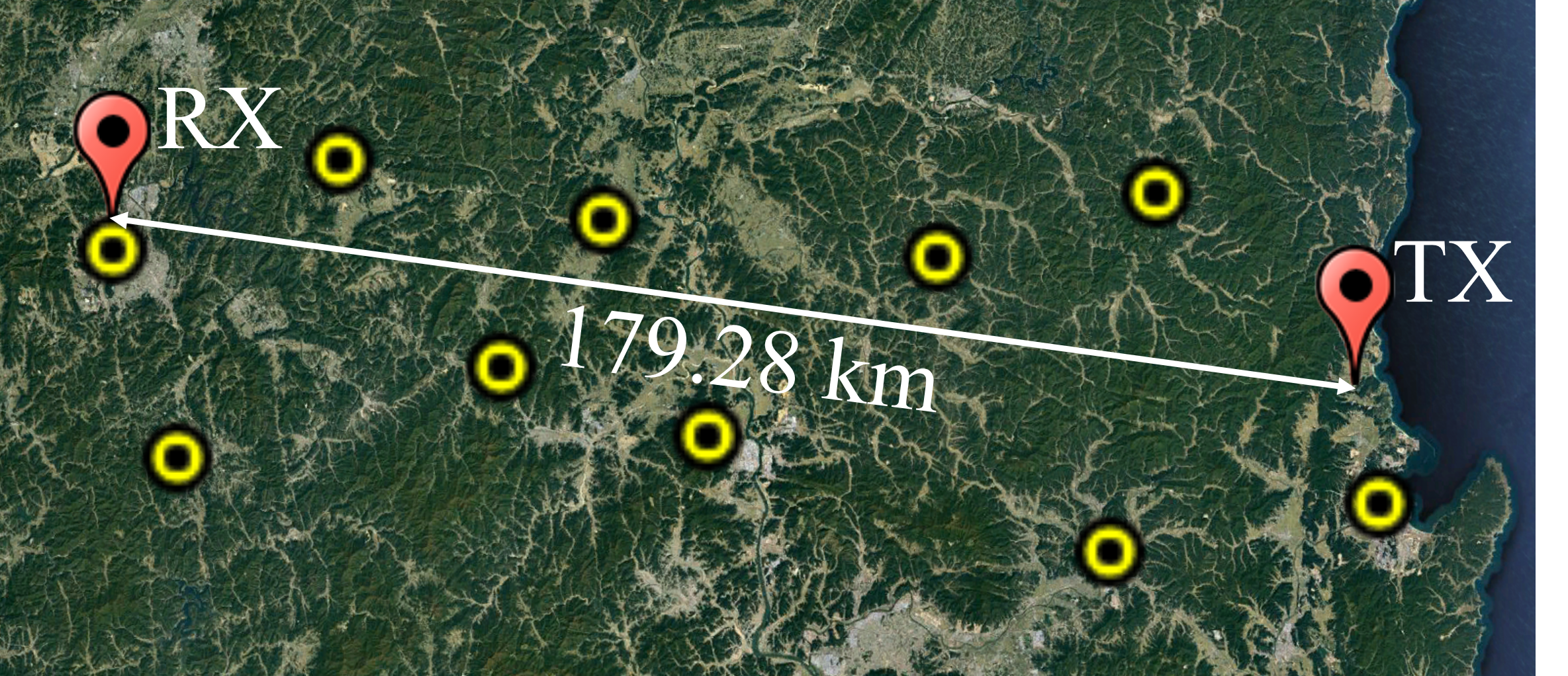}
  \caption{Locations of the Pohang eLoran transmitter (TX), eLoran timing receiver (RX) in Daejeon, and ten weather stations (indicated by yellow circles) near the groundwave propagation path.}
  \label{fig:SiteMap}
\end{figure}

This data collection setup was installed at the Korea Research Institute of Ships and Ocean Engineering (KRISO) in Daejeon, Korea. 
The received eLoran signals were transmitted from the Pohang transmitter (9930M). 
The distance between the Pohang transmitter and the receiver at KRISO is 179.28~km, as illustrated in Figure~\ref{fig:SiteMap}.

The meteorological data used to train our estimation models were obtained from the Korea Meteorological Administration (KMA). 
KMA provides various meteorological measurements recorded hourly at 102 weather stations nationwide. 
The geographic coordinates of these weather stations are also provided by KMA. 
The locations of the ten weather stations used in this study are indicated by yellow circles in Figure~\ref{fig:SiteMap}.

\subsection{Correlation Analysis Between eLoran/GPS TD and Meteorological Factors}
\label{sec:analysis}

\begin{figure*}
  \centering
  \includegraphics[width=0.7\linewidth]{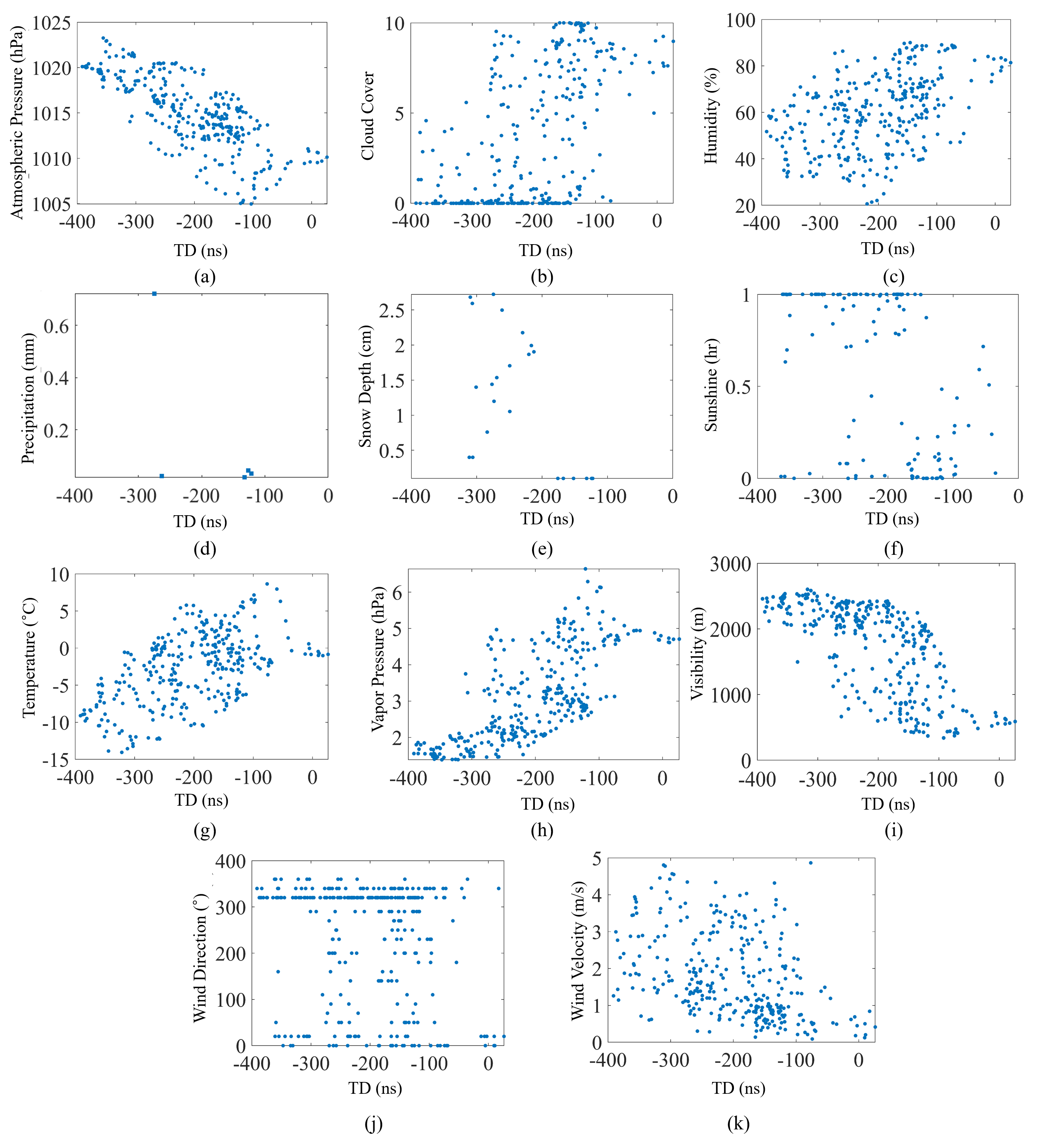}
  \caption{Relationship between eLoran/GPS TD measurements and eleven meteorological factors over a month at a static location: (a) atmospheric pressure, (b) cloud cover (represented by an integer number without any unit between 0 and 10), (c) humidity, (d) precipitation, (e) snow depth, (f) sunshine (represented by the duration within one-hour intervals that sunshine touches a specific location on the ground), (g) temperature, (h) vapor pressure, (i) visibility, (j) wind direction, and (k) wind velocity.
  }
  \label{fig:FeatureRelation}
\end{figure*}

Figure~\ref{fig:FeatureRelation} shows the relationships between eLoran/GPS TD measurements and eleven meteorological factors at the receiver location, based on one month of data. 
The nearest weather station to the receiver is located approximately 3~km away. 
For the plots in Figure~\ref{fig:FeatureRelation}, the meteorological data were interpolated to the grid point nearest to the receiver’s location using the interpolation method described in Section~\ref{sec:Interpolation}.

To evaluate the correlation between eLoran/GPS TD and meteorological factors, Pearson's correlation coefficients were calculated. 
Table~\ref{table:Pearson} presents the Pearson correlation coefficients between each of the eleven meteorological factors and the eLoran/GPS TD, based on the one-month dataset shown in Figure~\ref{fig:FeatureRelation}. 
Among the eleven meteorological factors, the $p$-value for precipitation exceeds 0.05, indicating that the correlation between precipitation and eLoran/GPS TD is statistically insignificant. 
Among the remaining ten factors, seven (atmospheric pressure, cloud cover, humidity, temperature, vapor pressure, visibility, and wind velocity) exhibit a strong correlation with eLoran/GPS TD, with the magnitude of Pearson's correlation coefficient $r$ being 0.5 or higher. 
The other three factors (snow depth, sunshine, and wind direction) show relatively lower $r$ values.

Based on this correlation analysis, we propose using seven meteorological factors---atmospheric pressure, cloud cover, humidity, temperature, vapor pressure, visibility, and wind velocity---to train our estimation models, which are detailed in Section~\ref{sec:EstModels}. 
Among these seven factors, three (atmospheric pressure, humidity, and temperature) were used in \cite{Pu2021:Accuracy}, and five (humidity, temperature, vapor pressure, visibility, and wind velocity) were used in \cite{Pu2019:Analysis}. 
In \cite{Liu2023:eLoran}, seven factors---atmospheric pressure, humidity, precipitation, temperature, vapor pressure, wind direction, and wind velocity---were used, partially overlapping with our selected set. 
Compared to \cite{Liu2023:eLoran}, we excluded two factors (precipitation and wind direction) that showed either statistical insignificance or low correlation, and added two factors (cloud cover and visibility). 
The benefits of incorporating these additional meteorological factors are evaluated in Section~\ref{sec:Results}.

\begin{table}
\renewcommand{\arraystretch}{1.3}
\caption{Pearson's Correlation Coefficient Between Each of the Eleven Meteorological Factors and eLoran/GPS TD Measurements}
\label{table:Pearson}
\centering
\begin{tabular}{w{c}{3.5cm} !{\vline} w{c}{1cm} w{c}{1.3cm} }
\thickhline
\textbf{Meteorological Factors} & $\bm{r}$ & $\bm{p}$ \\
\thickhline
Atmospheric Pressure [hPa] & $-0.72$ & $<$ 0.001\\
\hline
Cloud Cover [-] & $0.60$ & $<$ 0.001\\
\hline
Humidity [\%] & $0.51$ & $<$ 0.001\\
\hline
Precipitation [mm] & $-0.64$ & $>$ 0.05 \\
\hline
Snow Depth [cm] & $-0.46$ & $<$ 0.05\\
\hline
Sunshine [hr] & $-0.46$ & $<$ 0.001\\
\hline
Temperature [$^\circ$C] & $0.50$ & $<$ 0.001\\
\hline
Vapor Pressure [hPa] & $0.71$ & $<$ 0.001\\
\hline
Visibility [m] & $-0.68$ & $<$ 0.001\\
\hline
Wind Direction [$^\circ$] & $-0.17$ & $<$ 0.05 \\
\hline
Wind Velocity [m/s] & $-0.52$ & $<$ 0.001\\
\hline
\thickhline
\end{tabular}
\centering
\begin{tabular}{@{}p{\dimexpr\linewidth-2\tabcolsep}@{}}
\footnotesize Note: The Korea Meteorological Administration represents the cloud cover by an integer number without any unit between 0 and 10.
\end{tabular}
\end{table}

\subsection{Preparation of Meteorological Data and Terrain Elevation Data for eLoran/GPS TD Estimation}
\label{sec:Interpolation}

To acquire meteorological and terrain elevation data along the groundwave propagation path, meteorological data maps and a terrain elevation profile are generated. 
The meteorological data map is a grid map with a grid size of 0.01$^\circ$ in both latitude and longitude, covering the area between the transmitter and the receiver. 
For each meteorological factor and each hour, a separate meteorological data map is created. 
To construct these maps, collected meteorological data are assigned to the nearest grid points. 
Unassigned grid points are filled using inverse distance weighting (IDW) interpolation. 
IDW has been shown to be effective for generating high-resolution grid maps of various meteorological factors \cite{Jo2018:Applicability}.

IDW calculates interpolated values for unassigned grid points by computing the weighted average of the values at assigned grid points, using inverse distances as weights. 
Shepard's method \cite{Shepard1968:A} is employed to implement IDW. 
The interpolation function of Shepard's method, which determines an interpolated value $S({\mathbf{p}})$ at a two-dimensional grid point ${\mathbf{p}}$, is defined as follows:
\begin{equation}
\label{eqn:IDW} 
  S({\mathbf{p}})=
  \begin{cases}
    \frac{\sum_{i=1}^{N_g} w_i S({{\mathbf{p}}_i})}
         {\sum_{i=1}^{N_g} w_i}
    = \frac{\sum_{i=1}^{N_g}  \frac{S({{\mathbf{p}}_i})}{dist({\mathbf{p}}, {\mathbf{p}}_i)}} 
           {\sum_{i=1}^{N_g} \frac{1}{dist({\mathbf{p}}, {\mathbf{p}}_i)}}, 
    & \text{if}\ dist({\mathbf{p}},{\mathbf{p}}_i) \neq 0 \\
    S({{\mathbf{p}}_i}), 
    & \text{if}\ dist({\mathbf{p}},{\mathbf{p}}_i)=0 \\
  \end{cases}
\end{equation}
Here, $N_g$ is the total number of grid points assigned with measured meteorological data values, ${\mathbf p}_i$ denotes a grid point with an assigned value, and $dist({\mathbf p}, {\mathbf p}_i)$ represents the distance between ${\mathbf p}$ and ${\mathbf p}_i$. 
The weighting factor $w_i$ for IDW is defined as $\frac{1}{dist({\mathbf p}, {\mathbf p}_i)}$.

We observed that one of the main limitations of existing eLoran/GPS TD estimation methods in the literature is that, while they consider meteorological data at various points along the propagation path, they do not account for the topographical characteristics of each location. 
Given that the propagation delay of groundwaves is influenced by terrain elevation along the propagation path \cite{Zhou2011:LF, Zhou2021:Study, Wang2022:Method, Chang2023:Evaluation}, the eLoran/GPS TD can vary with changes in terrain elevation. 
Therefore, terrain elevation should also be considered as a feature for eLoran/GPS TD estimation, in addition to meteorological factors.

However, incorporating terrain elevation data as input nodes alongside time-varying meteorological data does not lead to performance improvements in the TD estimation model. 
Unlike meteorological factors, terrain elevation remains constant over time, meaning that only the input nodes corresponding to time-varying meteorological data contribute effectively to the learning process, while those with static terrain elevation data do not.

\begin{figure}
  \centering
  \includegraphics[width=0.9\linewidth]{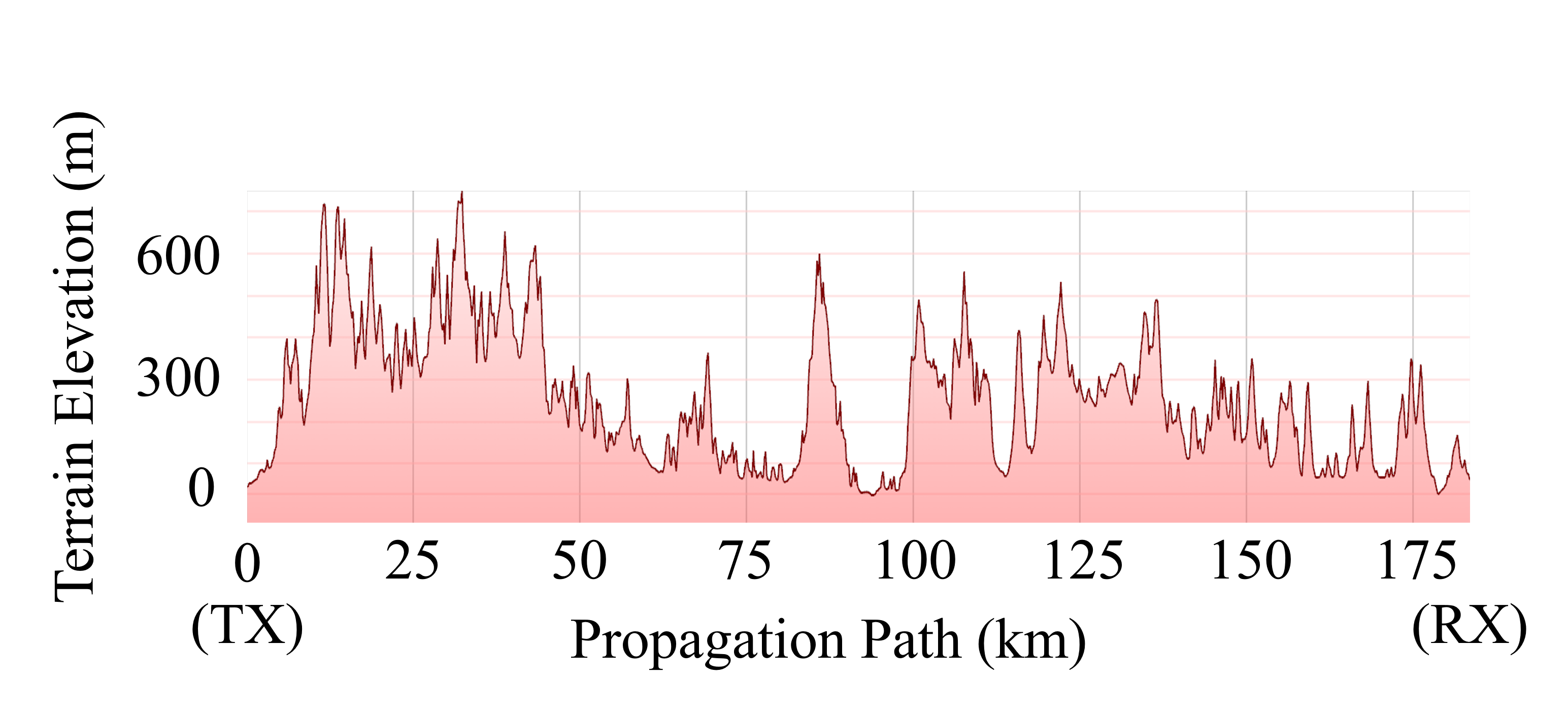}
  \caption{Terrain elevation profile between the Pohang transmitter (9930M) and the eLoran timing receiver in Daejeon.} 
  \label{fig:Elevation}
\end{figure}

Groundwaves take longer to traverse high-altitude terrain than low-altitude terrain. 
To enhance TD estimation performance, we weighted the meteorological data using terrain elevation values at each point and incorporated these weighted values into the learning process. 
The detailed weighting and learning scheme is described in Section~\ref{sec:WLR--AGRNN2}. 
The terrain elevation profile was generated using a digital elevation model (DEM) database \cite{GPSV, Farr2007:Shuttle} with a resolution of 30~meters. 
An example of the generated terrain elevation profile is shown in Figure~\ref{fig:Elevation}.

\section{eLoran/GPS TD Estimation Models}  
\label{sec:EstModels}

This section provides a detailed explanation of the two proposed eLoran/GPS TD estimation models.

\subsection{LASSO-Regularized MPR Model}

Multivariate polynomial regression (MPR), which has not previously been applied to eLoran/GPS TD estimation, is a form of regression analysis commonly used for nonlinear modeling, similar to backpropagation neural networks (BPNN) or general regression neural networks (GRNN). 
Unlike standard linear regression, polynomial regression computes the output as a sum of polynomial terms of the input variables up to a specified $m$-th degree. 
This allows it to model complex nonlinear relationships between inputs and outputs that conventional linear regression cannot capture.

If the MPR model \cite{Wei2016:Higher} is applied to eLoran/GPS TD estimation, the relationship between the estimated eLoran/GPS TD value $\hat{y}$ and the $m$-th order polynomial terms of $n$ meteorological factors at a single epoch can be expressed as:
\begin{equation}
\label{eqn:MultiPolyTerm}
\begin{split}
  \hat{y} = &~\beta_0 + \sum_{i_1=1}^{n} \beta_{i_1} x^{(i_1)} + \sum_{i_1=1}^{n}\sum_{i_2=i_1}^{n} \beta_{i_1 i_2} x^{(i_1)} x^{(i_2)} + \dots + \\
  & \sum_{i_1=1}^{n}\sum_{i_2=i_1}^{n} \dots \sum_{i_k=i_{k-1}}^{n} \beta_{i_1 i_2 \dots i_k} x^{(i_1)} x^{(i_2)} \dots x^{(i_k)} + \dots +\\
  & \sum_{i_1=1}^{n}\sum_{i_2=i_1}^{n} \dots \sum_{i_m=i_{m-1}}^{n} \beta_{i_1 i_2 \dots i_m} x^{(i_1)} x^{(i_2)} \dots x^{(i_m)}
\end{split}
\end{equation}
Here, $x^{(i_k)}$ ($1 \leq i_k \leq n,\ 1 \leq k \leq m$) is the $i_k$-th input variable (i.e., the $i_k$-th meteorological factor) used to estimate $\hat{y}$ (the eLoran/GPS TD value), and $\beta_{i_1 i_2 \dots i_k}$ is the weight assigned to each polynomial term $x^{(i_1)} x^{(i_2)} \dots x^{(i_k)}$. 
An interaction term refers to the product of two or more input variables in the regression model, such as $x^{(i_1)} x^{(i_2)}$ or $x^{(i_1)} x^{(i_2)} \dots x^{(i_k)}$.

\begin{figure}
  \centering
  \includegraphics[width=0.7\linewidth]{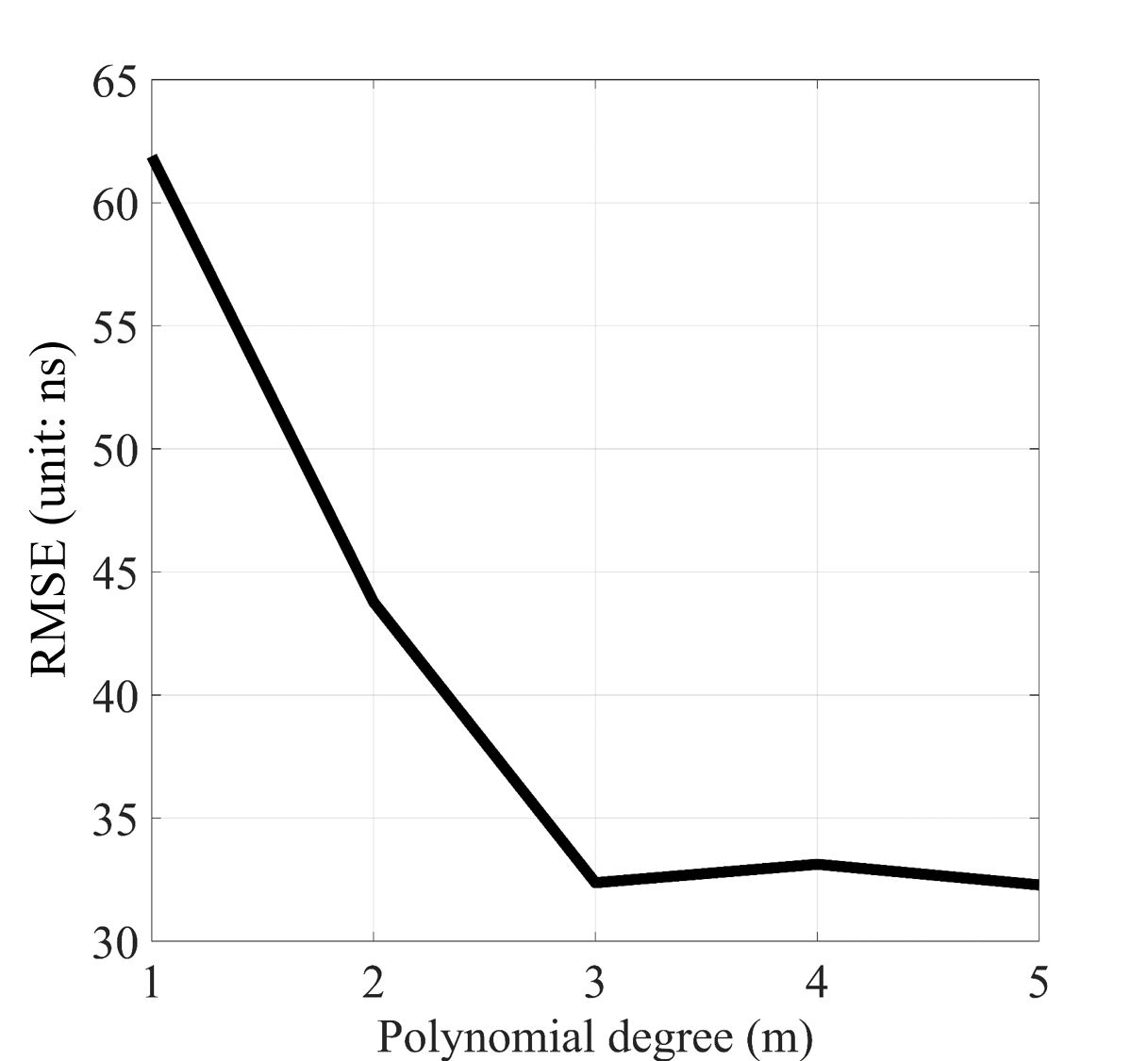}
  \caption{Root mean square error (RMSE) versus polynomial degree for the LASSO-regularized MPR model. The optimal degree is selected as $m = 3$, where RMSE convergence is observed.}
  \label{fig:polyDeg}
\end{figure}

The meteorological factors used in eLoran/GPS TD estimation are not independent but exhibit complex correlations. 
For example, the interdependencies among pressure, temperature, and humidity have been studied in \cite{Beniston2002:Shifts}. 
The estimation model we aim to develop also includes meteorological factors with evident correlations, such as visibility and cloud cover. 
The interaction terms in MPR can effectively capture these combined influences in the estimation process \cite{Li2020:Deeply}. 
For these reasons, we adopt the MPR method with third-order polynomial terms as the highest-order terms (i.e., $m = 3$). 
The RMSE of the model for polynomial degrees $m$ ranging from 1 to 5 is presented in Figure~\ref{fig:polyDeg}. 
As shown in the figure, the RMSE appears to converge after $m = 3$; therefore, we set $m = 3$ for our model.

To improve the accuracy of polynomial regression, it is often necessary to increase the number of input variables and the degree of the model, which results in a larger number of parameters to estimate. 
However, in cases involving multiple input variables, as in our study, this can lead to highly unstable models \cite{Peckov2012:A, Su2021:Regularized}. 
To address this issue, Su \textit{et al.} \cite{Su2021:Regularized} proposed applying the least absolute shrinkage and selection operator (LASSO) regression to the MPR model. 
LASSO regression constrains the weights of polynomial terms, making the model more robust to problems such as overfitting.

The goal of LASSO regression \cite{Su2021:Regularized} is to minimize the following loss function:
\begin{equation}
\label{eqn:Lasso2}
  \sum_{t=1}^{T} (y_t-\hat{y}_t)^2+\alpha \sum_{p=1}^{P} |w_p|
\end{equation}
where $T$ is the number of epochs, $y_t$ is the measured eLoran/GPS TD value at the $t$-th epoch, $\hat{y}_t$ is the estimated eLoran/GPS TD value at the $t$-th epoch, $P$ is the number of polynomial terms in the MPR model (i.e., $P = \binom{n}{1} + \binom{n+1}{2} + \dots + \binom{n+m-1}{m}$), $w_p$ is the weight of the $p$-th polynomial term, and $\alpha$ is the LASSO regularization parameter that controls the amount of shrinkage.

\begin{figure}
  \centering
  \includegraphics[width=0.8\linewidth]{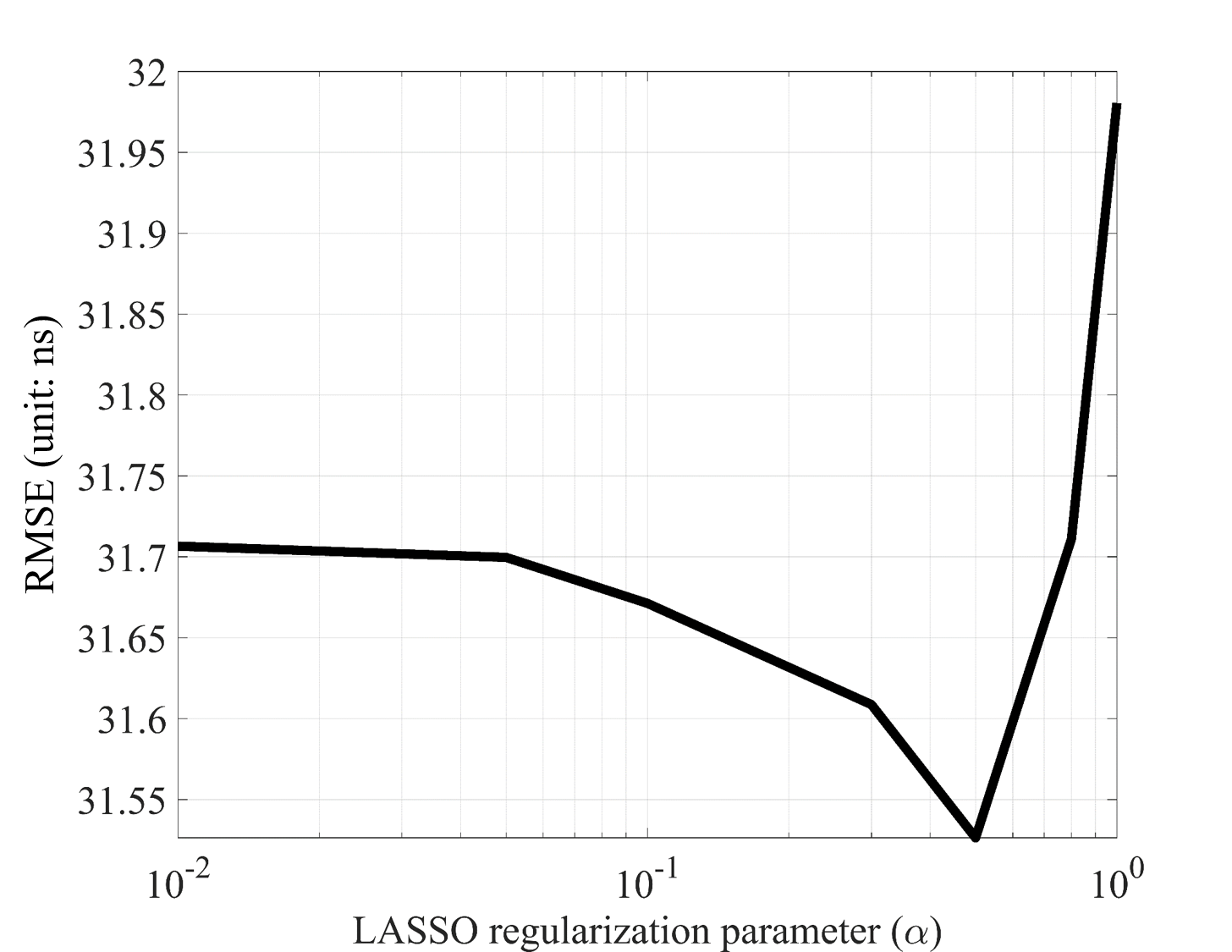}
  \caption{Sensitivity analysis of the LASSO regularization parameter $\alpha$. The RMSE of eLoran/GPS TD estimation is plotted as a function of $\alpha$ on a logarithmic scale, revealing that the optimal value is approximately $\alpha = 0.5$.}
  \label{fig:SensitivityAnalysis}
\end{figure}

To find the optimal value of $\alpha$ for our model, we conducted a sensitivity analysis to evaluate its impact on predictive performance.
Figure~\ref{fig:SensitivityAnalysis} shows the variation in eLoran/GPS TD estimation error, measured by root mean square error (RMSE), as $\alpha$ changes on a logarithmic scale.
As shown in the figure, RMSE gradually decreases as $\alpha$ increases from very small values, reaching a minimum at approximately $\alpha = 0.5$.
This behavior aligns with theoretical expectations of LASSO regularization: at lower values of $\alpha$, the model may experience overfitting due to insufficient regularization, resulting in higher errors \cite{Emmert2019:High, Huri2017:Selecting}.
Beyond the optimal value ($\alpha = 0.5$), further increases in $\alpha$ lead to excessive regularization, resulting in underfitting and subsequently higher estimation errors.
Therefore, based on this sensitivity analysis, we determined the optimal regularization parameter for balancing model complexity and predictive performance to be $\alpha = 0.5$.

\begin{figure}
  \centering
  \includegraphics[width=1.0\linewidth]{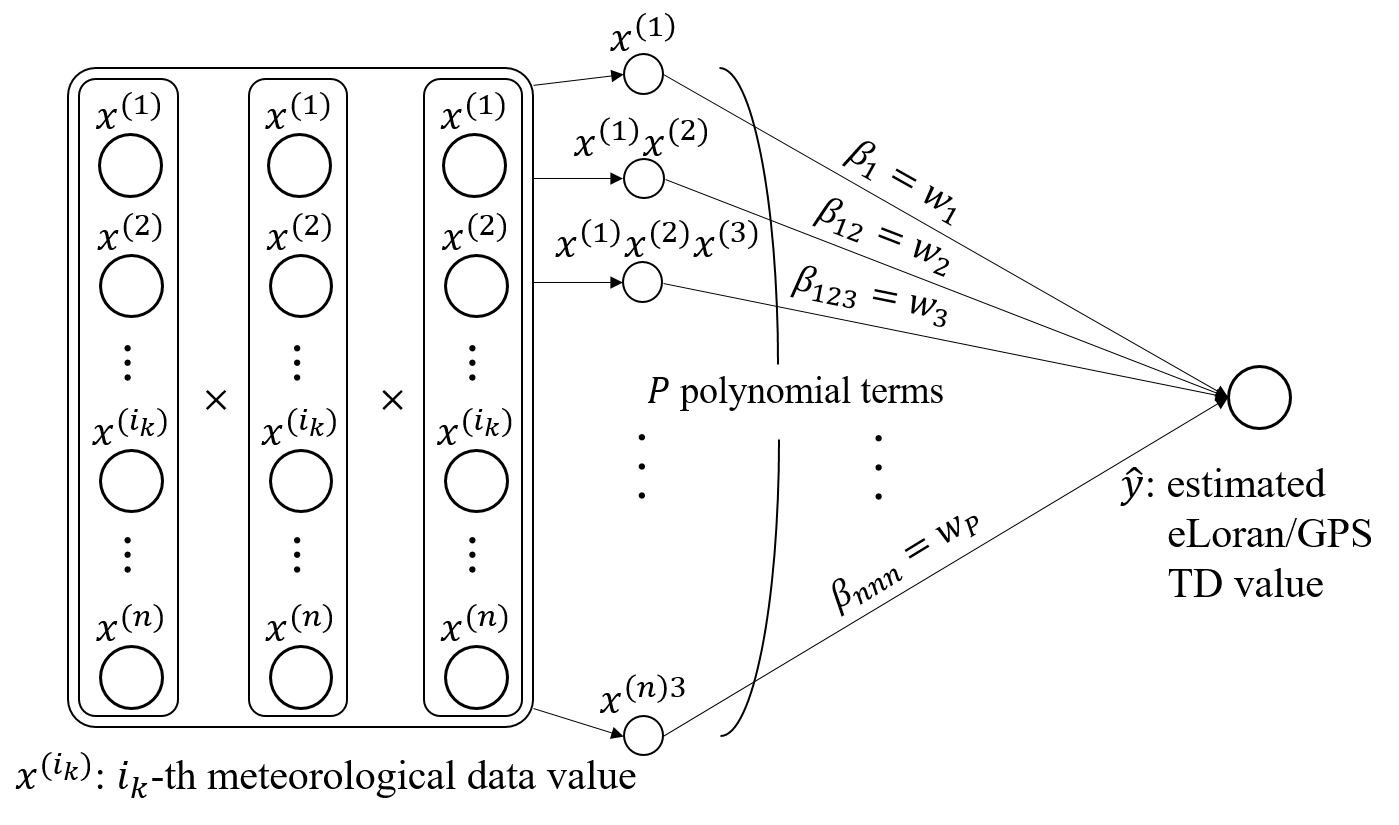}
  \caption{Structure of the LASSO-regularized third-order MPR model for GPS/eLoran TD estimation.}
  \label{fig:PolyLasso}
\end{figure}

LASSO adds an L1 regularization term, $\alpha \sum_{p=1}^{P} |w_p|$, to the least squares loss function used in conventional MPR to adjust the weights. 
This term penalizes polynomial terms whose weights are close to zero and do not significantly contribute to the learning process. 
As a result, the learning process becomes more efficient without compromising accuracy and is also more resistant to overfitting \cite{Su2021:Regularized}. 
The structure and algorithm of our eLoran/GPS TD estimation based on the ``LASSO-regularized MPR model'' \cite{Su2021:Regularized} are shown in Figure~\ref{fig:PolyLasso} and Algorithm~\ref{alg:PolyLasso}, respectively. 
To the best of our knowledge, this is the first adaptation of the LASSO-regularized MPR model to the eLoran/GPS TD estimation problem.

\begin{algorithm}
  \caption{Training procedure for the LASSO-regularized MPR model used for eLoran/GPS TD estimation.}
  \label{alg:PolyLasso}
  \begin{algorithmic}[1]
    \Require A training dataset of $T$ epochs, each containing meteorological data, $\mathbf{x}_t = \left[\begin{array}{lll} x_t^{(1)} & \dots & x_t^{(n)} \end{array}\right]^T$, and measured eLoran/GPS TD, $y_t$ $(t = 1, \dots, T)$
    \Ensure Estimated TD, $\hat{y}$
    \State Initialize network weights $w_p$
    \While{Iteration}
        \For{$t = 1$ to $T$}
            \State Estimate $\hat{y}_t$ using (\ref{eqn:MultiPolyTerm}) for $\mathbf{x}_t$
        \EndFor
        \State Update network weights $w_p$ by minimizing the loss function in (\ref{eqn:Lasso2}) using $y_t$ and $\hat{y}_t$
    \EndWhile
    \State Using the updated network weights $w_p$, estimate $\hat{y}$ for newly collected meteorological data, $\mathbf{x}$
  \end{algorithmic}
\end{algorithm}

The computational complexity of the LASSO-regularized MPR model is dominated by the generation of polynomial features. 
Given the dataset size $N$, the number of meteorological factors $n$, and the polynomial order $m$, the complexity of generating polynomial features can be expressed as:
\begin{equation}
    O(N \times n^{m})
\end{equation}
This complexity arises from the combinatorial explosion of features generated when combining the original meteorological variables up to the $m$-th order. 
As a result, the computational burden increases rapidly as $n$ and $m$ grow.

\subsection{WLR--AGRNN Model}
\label{sec:WLR--AGRNN}

\subsubsection{Existing Methods and Limitations}

Previous studies \cite{Pu2021:Accuracy, Liu2023:eLoran} have demonstrated that utilizing meteorological data collected at multiple locations along the propagation path---rather than solely at the receiver's location---can improve the prediction accuracy of eLoran/GPS TD. 
However, these existing approaches rely on data measured at sparse, discrete locations, which limits the model’s ability to capture localized environmental variations along the path.

To overcome this limitation, we propose a denser sampling of meteorological data points by interpolating data along the signal propagation path, as described in Section~\ref{sec:Interpolation}. 
Through this interpolation, the number of input nodes in the network increases, with each interpolated data point serving as a separate input node. 
This approach, however, introduces a new challenge: neighboring nodes often exhibit highly similar values due to spatial proximity and interpolation. 
Such redundancy may cause the learning process to prioritize repeated values rather than capturing the spatial variations essential for accurate TD estimation. 
To mitigate this issue, a mechanism is required to regulate the influence of each input node, preventing the dominance of similar values and ensuring that the network maintains focus on the relationship between each location and the eLoran/GPS TD.

We address this by constructing a composite network consisting of specialized subnetworks for each input data type. 
When training on datasets comprising heterogeneous data types with distinct characteristics, employing separate subnetworks for each type---while combining their outputs---has been shown to yield superior learning performance compared to unified network architectures \cite{Jacobs1991:Adaptive}. 
This design enables the network to account for the unique attributes of each data type, ultimately improving overall performance.

A representative example of this composite network structure is the Mixture of Experts (MoE). 
The simplest form of MoE \cite{Jacobs1991:Adaptive} employs feedforward neural networks as the subnetworks (expert networks), with the final output generated through a weighted summation of the outputs from each subnetwork. 
Although this approach has not yet been applied to eLoran/GPS TD estimation, it can be implemented by designing the network structure similarly to how BPNN or GRNN have been applied in previous TD estimation studies.

In this work, we aim to develop a network better suited for TD estimation by modifying both the expert networks and the output aggregation method compared to the basic MoE model. 
A detailed explanation of the proposed network is provided in the following subsection.

\subsubsection{Proposed Method}
\label{sec:WLR--AGRNN2}

As shown in Section~\ref{sec:analysis}, the meteorological factors selected for our eLoran/GPS TD estimation model exhibit both linear and nonlinear relationships with eLoran/GPS TD. 
To capture these relationships, we designed a composite network composed of two distinct components: one for the expert network and the other for final output aggregation. 
We selected WLR as the expert network to model individual meteorological factors and AGRNN as the output aggregation network to produce the final TD estimate. 
The rationale behind this design is that WLR effectively captures the linear dependencies between meteorological factors and eLoran/GPS TD, while AGRNN is well-suited to model the nonlinear interactions among the outputs of the expert networks.

First, compared to the feedforward network used as the expert network in the basic MoE model \cite{Jacobs1991:Adaptive}, WLR is more suitable for our case due to the large volume of data. 
In ordinary linear regression, the residual sum of squares (RSS) is minimized, where all data points contribute equally:
\begin{equation}
    RS\!S = \sum_{t=1}^{T} (y_t - \hat{y}_t)^2
\end{equation}
In contrast, WLR minimizes the weighted residual sum of squares (WRSS), which assigns different weights to each data point:
\begin{equation}
\label{eqn:WRSS}
    WRS\!S = \sum_{t=1}^{T} w_{t} (y_t - \hat{y}_t)^2
\end{equation}
where $w_t$ denotes the weight applied to the $t$-th epoch. 
While weighting is beneficial regardless of dataset size, its advantages become more pronounced as the number of input nodes increases. 
With larger datasets, the relative contribution of individual data points diminishes, and critical data may become underrepresented during the learning process. 
By dynamically adjusting the importance of each data point, WLR ensures that significant data points have a stronger influence on the model, thereby optimizing the learning process \cite{Wu2018:Evaluation}.
This is particularly valuable in our study, where the large number of input nodes necessitates an efficient mechanism to manage and prioritize data contributions.

Second, AGRNN---like the GRNN used in a previous eLoran/GPS TD estimation study \cite{Pu2021:Accuracy}---is well-suited for capturing nonlinear relationships among the expert network outputs. 
When combined with WLR's ability to model linear relationships, this architecture is expected to outperform the basic MoE model's weighted averaging approach.

GRNN can estimate the eLoran/GPS TD value, $\hat{y}$, using the following equation \cite{Pu2021:Accuracy}:
\begin{equation}
\label{eqn:GRNN}
    \hat{y} = \frac{\sum_{t=1}^{T} y_t \exp\left(-\frac{\|\mathbf{\hat{x}} - \mathbf{\hat{x}}_t\|^2}{2\sigma^2}\right)}{\sum_{t=1}^{T} \exp\left(-\frac{\|\mathbf{\hat{x}} - \mathbf{\hat{x}}_t\|^2}{2\sigma^2}\right)}
\end{equation}
where $T$ is the number of epochs in the training data, $y_t$ is the measured eLoran/GPS TD value at the $t$-th epoch in the training data, $\mathbf{\hat{x}}$ is the input data vector (i.e., meteorological factors) at an epoch in the test data used to estimate $\hat{y}$ for the same epoch, $\mathbf{\hat{x}}_t$ is the input data vector at the $t$-th epoch in the training data, and $\sigma$ is a smoothing factor.
A conventional GRNN uses the same smoothing factor, $\sigma$, for all features. 
This uniformity can limit the model's ability to accurately capture local characteristics and complexities in the input data \cite{Specht1994:Experience, Amato2020:On}. 
Furthermore, a fixed smoothing parameter optimized for one dataset may perform poorly on others due to its lack of adaptability.

AGRNN \cite{Specht1994:Experience, Subashini2024:Smart} differs from conventional GRNN by utilizing an anisotropic Gaussian kernel function, which assigns a distinct smoothing factor to each feature.
Instead of applying a single smoothing factor $\sigma$, as in GRNN, AGRNN assigns a smoothing factor $\sigma_i$ ($i = 1, 2, \dots, n$) to each feature in the dataset. 
Accordingly, AGRNN estimates the output $\hat{y}$ using the following equation \cite{Amato2020:On}:
\begin{equation}
\label{eqn:AGRNN}
    \hat{y} = \hat{y}(\hat{\mathbf{x}}, \mathbf{U}) = \frac{\sum_{t=1}^{T} y_t \exp \left( -\sum_{i=1}^n \frac{(\hat{x}^{(i)} - \hat{x}_t^{(i)})^2}{2\sigma_i^2} \right)}{\sum_{t=1}^{T} \exp \left( -\sum_{i=1}^n \frac{(\hat{x}^{(i)} - \hat{x}_t^{(i)})^2}{2\sigma_i^2} \right)}
\end{equation}
Here, $\hat{x}^{(i)}$ is the $i$-th feature (i.e., the $i$-th meteorological factor) of the input data vector $\mathbf{\hat{x}}$ at an epoch in the test data used to estimate $\hat{y}$ for the same epoch, and $\hat{x}_t^{(i)}$ is the $i$-th feature of the input data vector $\mathbf{\hat{x}}_t$ at the $t$-th epoch in the training data. 
The matrix $\mathbf{U} = \left[\begin{array}{lll} \hat{\mathbf{x}}_1 & \dots & \hat{\mathbf{x}}_T \end{array}\right]_{n \times T}$ represents the full set of training input vectors.

In the AGRNN component, a distinct smoothing factor $\sigma_i$ is assigned to each feature to account for differences in scale and distribution. 
The selection method for $\sigma_i$ follows the approach introduced in \cite{Amato2020:On}, where feature-wise smoothing parameters are optimized based on kernel function behavior and local density considerations. 
This method has been shown to improve generalization in anisotropic kernel models, and we adopt it here to ensure that each meteorological feature contributes appropriately to the final TD estimation.

Although AGRNN has not previously been applied to eLoran/GPS TD estimation, its ability to account for the scale and distribution of each feature more comprehensively than GRNN makes it particularly well-suited for this task. 
Since eLoran/GPS TD estimation involves diverse meteorological factors with varying scales and distributions, AGRNN is well-equipped to model these complexities effectively.

To the best of the authors' knowledge, there are no existing examples in the literature where WLR and AGRNN have been combined as a composite network---neither for eLoran/GPS TD estimation nor for other neural network applications. 
Since WLR and AGRNN update their parameters differently, we modified the backpropagation process to accommodate both networks. 
After backpropagation adjusts the parameters, WLR produces an intermediate result. 
AGRNN then recalculates its required parameters, updated through backpropagation, to generate the final output.

\begin{figure}
  \centering
  \includegraphics[width=0.9\linewidth]{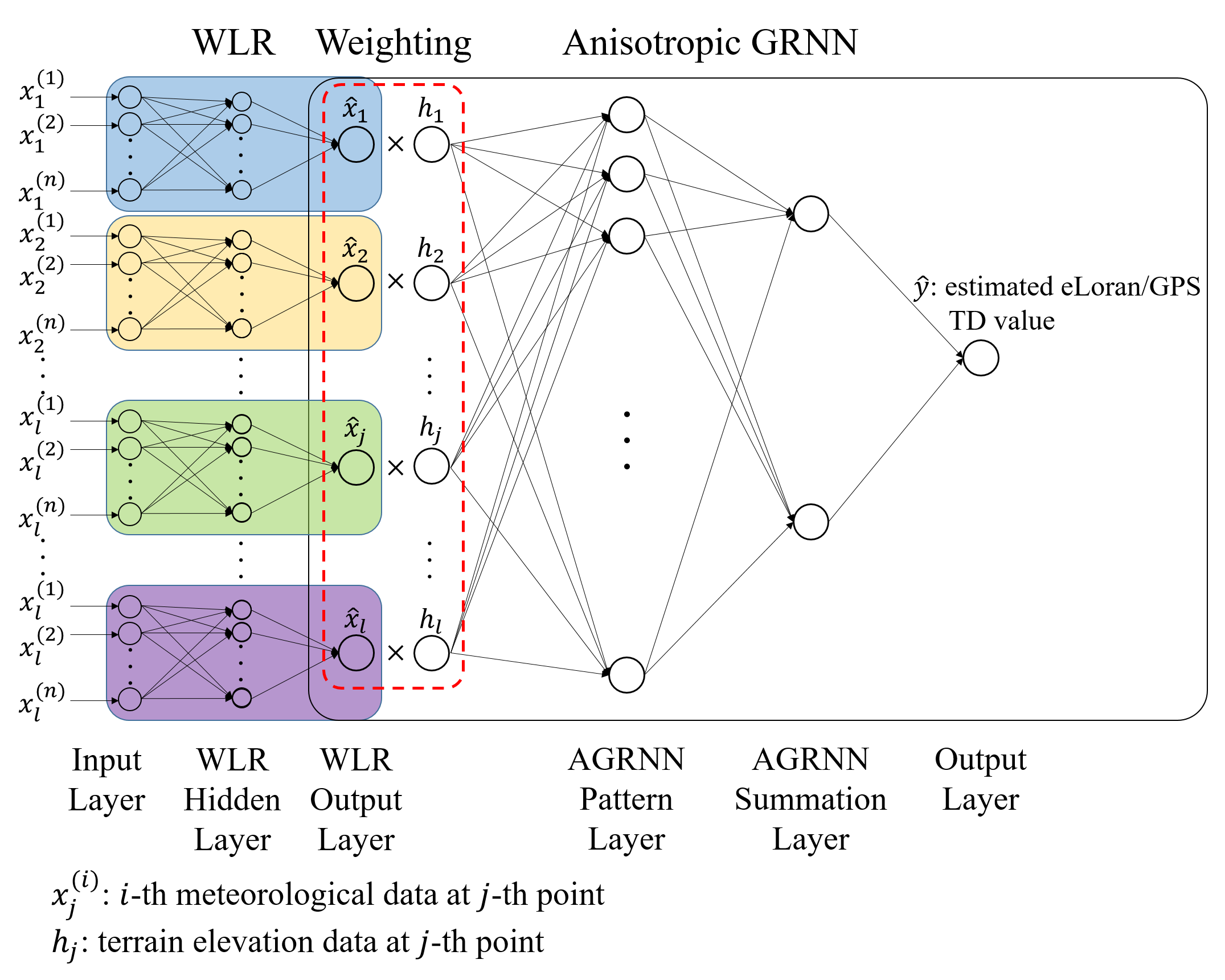}
  \caption{Architecture of the proposed WLR--AGRNN model for eLoran/GPS TD estimation, illustrating the integration of weighted linear regression (WLR) and anisotropic general regression neural network (AGRNN) for modeling linear and nonlinear relationships, respectively.}
  \label{fig:WLR--AGRNN}
\end{figure}

\begin{algorithm}
  \caption{Training procedure of the proposed WLR--AGRNN model for eLoran/GPS TD estimation.}
  \label{alg:WLR--AGRNN}
  \begin{algorithmic}[1]
    \Require Training dataset of $T$ epochs, each containing meteorological data $\mathbf{x}_{j,t} = \left[\begin{array}{lll} x^{(1)}_{j,t} & \dots & x^{(n)}_{j,t} \end{array}\right]^T$ $(j = 1, 2, \dots, l)$, terrain elevation $h_j$, and measured TD $y_t$ $(t = 1, \dots, T)$
    \Ensure Estimated TD, $\hat{y}$
    \State Initialize all network weights and biases
    \While{Iteration}
        \For{$t = 1$ to $T$}
            \For{$j = 1$ to $l$}
                \State Estimate $\hat{x}_{j,t}$ using (\ref{eqn:WLR1}) and (\ref{eqn:WLR2}) for $\mathbf{x}_{j,t}$
            \EndFor
            \State $\hat{\mathbf{x}}_{t,\mathrm{weighted}} =\left[\begin{array}{lll} \hat{x}_{1,t}h_1 & \dots & \hat{x}_{l,t}h_l \end{array}\right]^T$ 
        \EndFor
        \State $\mathbf{U} =\left[\begin{array}{lllll} \hat{\mathbf{x}}_{1,\mathrm{weighted}} & \dots & \hat{\mathbf{x}}_{T,\mathrm{weighted}} \end{array}\right]_{n \times T}$
        \For{$t = 1$ to $T$}
            \State $\hat{y}_t = \hat{y}(\hat{\mathbf{x}}_t, \mathbf{U})$ using (\ref{eqn:AGRNN})
        \EndFor
        \State Calculate the gradients of WRSS using (\ref{eqn:WRSS})
        \State Update $\mathbf{W}^{(k)}$ and $\mathbf{b}^{(k)}$ $(k=1,2)$ using Adam optimizer
    \EndWhile
    \State Estimate $\hat{y}$ for newly collected meteorological data $\mathbf{x}^{(i)}$ $(i = 1, 2, \dots, n)$ using updated parameters
    \end{algorithmic}
\end{algorithm}

The structure and algorithm of our WLR--AGRNN model are shown in Figure~\ref{fig:WLR--AGRNN} and Algorithm~\ref{alg:WLR--AGRNN}, respectively.

The implementation details of the WLR--AGRNN model are as follows:
\begin{itemize}
    \item The input layer processes the input data $\mathbf{x}_{1,t}$, $\mathbf{x}_{2,t}$, $\dots$, $\mathbf{x}_{j,t}$, $\dots$, $\mathbf{x}_{l,t}$ at the $t$-th epoch ($t = 1, 2, \dots, T$). 
    Each $\mathbf{x}_{j,t}$ $(j = 1, 2, \dots, l)$ represents the meteorological data collected at the $j$-th location and is defined as:
    \begin{equation}
    \label{eqn:weighted_data}
        \mathbf{x}_{j,t} =
        \begin{bmatrix}
            x^{(1)}_{j,t} \\
            x^{(2)}_{j,t} \\
            \vdots \\
            x^{(i)}_{j,t} \\
            \vdots \\
            x^{(n)}_{j,t}
        \end{bmatrix}
    \end{equation}
    where $n$ is the number of meteorological factors used in training, and $l$ is the number of locations along the signal propagation path from the eLoran transmitter to the receiver at which meteorological and terrain elevation data are available. 
    Here, $x^{(i)}_{j,t}$ denotes the $i$-th meteorological factor measured at the $j$-th location at the $t$-th epoch.

    \item Each $\mathbf{x}_{j,t}$ is used to estimate $\hat{x}_{j,t}$, a scalar representing the combined effects of meteorological factors at the $j$-th location. 
    The value $\hat{x}_{j,t}$ is computed as follows:
    \begin{equation}
    \label{eqn:WLR1}
        \mathbf{h}_{j,t} = \mathbf{W}^{(1)} \mathbf{x}_{j,t} + \mathbf{b}^{(1)}
    \end{equation}
    \begin{equation}
    \label{eqn:WLR2}
        \hat{x}_{j,t} = \mathbf{W}^{(2)} \mathbf{h}_{j,t} + \mathbf{b}^{(2)}
    \end{equation}
    where $\mathbf{h}_{j,t}$ is the output of the first hidden layer for the $j$-th location; $\mathbf{W}^{(1)}$ and $\mathbf{W}^{(2)}$ are the weight matrices for the WLR hidden and output layers, respectively; and $\mathbf{b}^{(1)}$ and $\mathbf{b}^{(2)}$ are the corresponding bias vectors.
    In terms of computational complexity, each of the $j$ measurement points, containing $n$ meteorological features, is independently transformed into a scalar, resulting in a complexity of $O(j \times n)$.

    \item Each $\hat{x}_{j,t}$ is weighted by $h_j$, the terrain elevation at the $j$-th location. 
    Inspired by the attention mechanism used in LSTM-based models, as described by Liu \textit{et al.} \cite{Liu2022:Machine}, this weighting approach enables the model to dynamically emphasize meteorological data from locations where elevation has a greater influence on eLoran/GPS timing delay. 
    While Liu \textit{et al.} \cite{Liu2022:Machine} used static features to learn attention weights and computed a weighted sum as the final LSTM state, we instead use the static feature (i.e., terrain elevation $h_j$) directly as the attention weight for each $\hat{x}_{j,t}$, as depicted in Figure~\ref{fig:WeightingStructure}. 
    
    In Liu \textit{et al.} \cite{Liu2022:Machine}, multiple dynamic and static features are considered for each measurement location, making it logical to learn separate attention weights for each feature by incorporating both dynamic and static information. 
    In contrast, our model involves only a single static feature per measurement location. 
    Thus, it is more efficient and practical to first aggregate the dynamic features through learning and then directly apply the static feature as an attention weight, allowing effective modulation during backpropagation.

    It is important to note that, unlike conventional attention mechanisms commonly used in deep learning---where attention weights are dynamically learned from the data---the proposed weighting approach utilizes terrain elevation as a fixed, predetermined weight.
    
    During training, when $\hat{x}_{j,t}$ is updated, $h_j$ ensures that the output from the previous WLR step is modulated according to the terrain information. 
    This facilitates faster and more accurate model learning, allowing the network to dynamically highlight measurement points that are more significantly impacted by elevation changes, thereby capturing nonlinear interactions more effectively than a simple linear combination. 
    Since these operations scale linearly with the number of measurement points, the overall complexity of this step is $O(j)$.

    \begin{figure}
    \centering
    \includegraphics[width=0.9\linewidth]{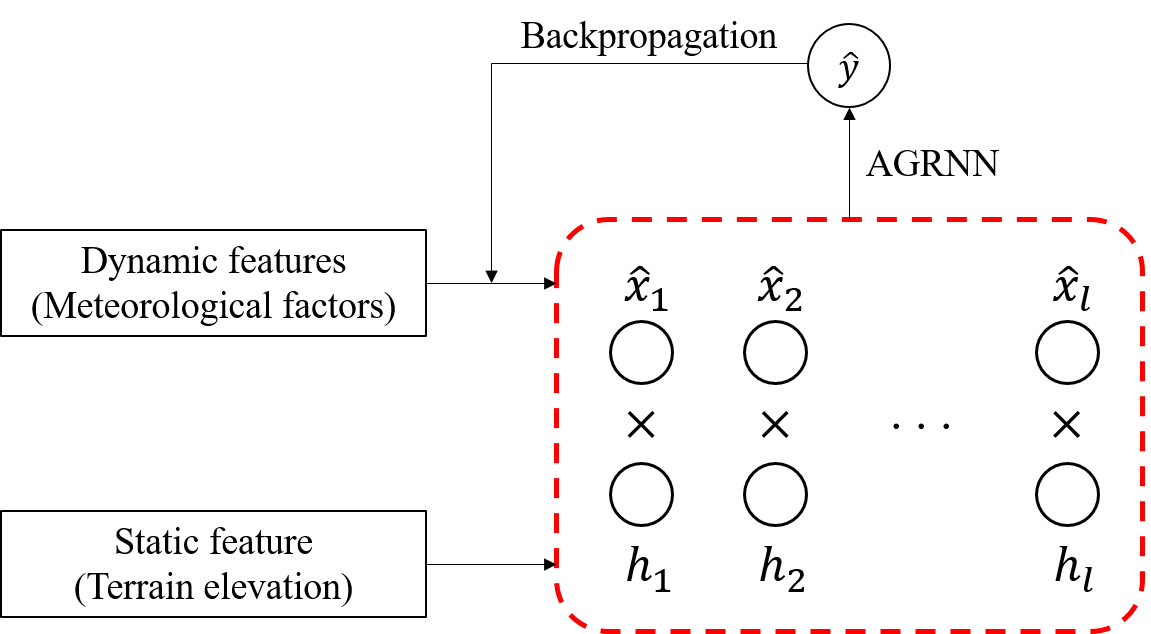}
        \caption{Weighting structure of the proposed WLR--AGRNN model. 
        Each expert network output $\hat{x}_j$ (derived from dynamic meteorological features) is weighted by the corresponding static terrain elevation $h_j$ before being passed to the AGRNN. 
        During training, backpropagation updates the parameters of the expert networks, while the AGRNN aggregates the weighted outputs to estimate the final eLoran/GPS TD.}
        \label{fig:WeightingStructure}
    \end{figure}

    \item The eLoran/GPS TD value $\hat{y}_t$ is estimated using AGRNN, as described in (\ref{eqn:AGRNN}). 
    AGRNN performs kernel-based computations over the entire training dataset of $N$ samples, resulting in a computational complexity of $O(N)$.

    \item After AGRNN estimates $\hat{y}_t$ for all epochs in the dataset, backpropagation is performed to train the WLR network.
    The gradients of the weighted residual sum of squares (WRSS) are computed during the backpropagation process, as defined in (\ref{eqn:WRSS}). 
    The Adam optimizer is used to update the WLR weights and biases, $\mathbf{W}^{(k)}$ and $\mathbf{b}^{(k)}$ ($k = 1, 2$), due to its superior convergence performance compared to standard gradient descent \cite{Kingma2014:Adam}. 
    By iterating until convergence, the model is trained to accurately estimate $\hat{y}$.
\end{itemize}

Considering all the above steps, the total computational complexity of the WLR--AGRNN model is $O(j \times n + j + N)$. 
Since the AGRNN module typically dominates the overall computation, the final complexity is effectively $O(N)$. 
Therefore, the WLR--AGRNN model demonstrates better computational efficiency than the LASSO-regularized MPR model, particularly in scenarios involving large feature spaces and higher-order polynomial terms.

\section{Experimental Results} 
\label{sec:Results}

\begin{table*}
\renewcommand{\arraystretch}{1.2}
\caption{RMSEs of eLoran/GPS TD Estimation.}
\label{table:RMSE}
\centering
\begin{tabular}{w{c}{5cm} w{c}{1.2cm} w{c}{1.2cm} w{c}{1.2cm} w{c}{1.2cm} w{c}{1.2cm} w{c}{1.2cm}}
\thickhline
& \multicolumn{6}{c}{Locations where meteorological data were collected} \\ \cline{2-7}
& \multicolumn{3}{c}{Receiver location only} & \multicolumn{3}{|c}{10 locations near the propagation path} \\ \cline{2-7}
& \multicolumn{6}{c}{Number of meteorological factors applied} \\ \cline{2-7}
& 3 & 5 \cite{Pu2019:Analysis} & \textbf{7} & 3 \cite{Pu2021:Accuracy} & 5 & \textbf{7} \\ \cline{2-7}
\hline
BPNN \cite{Pu2019:Analysis, Liu2023:eLoran} [ns] & \multicolumn{1}{|c}{62} & 58 & 48 & \multicolumn{1}{|c}{60} & 53 & 45 \\
GRNN \cite{Pu2021:Accuracy} [ns] & \multicolumn{1}{|c}{56} & 55 & 45 & \multicolumn{1}{|c}{47} & 47 & 44 \\
MoE [ns] & \multicolumn{1}{|c}{48} & 42 & 31 & \multicolumn{1}{|c}{41} & 37 & 28 \\
\textbf{LASSO-regularized MPR [ns]} & \multicolumn{1}{|c}{50} & 43 & \textbf{35} & \multicolumn{1}{|c}{40} & 38 & \textbf{32} \\
\textbf{WLR--AGRNN [ns]} & \multicolumn{1}{|c}{38} & 35 & \textbf{30} & \multicolumn{1}{|c}{28} & 25 & \textbf{21} \\
\thickhline
\end{tabular}
\end{table*}

\begin{table*}
\renewcommand{\arraystretch}{1.2}
\caption{MAEs of eLoran/GPS TD Estimation.}
\label{table:MAE}
\centering
\begin{tabular}{w{c}{5cm} w{c}{1.2cm} w{c}{1.2cm} w{c}{1.2cm} w{c}{1.2cm} w{c}{1.2cm} w{c}{1.2cm}}
\thickhline
& \multicolumn{6}{c}{Locations where meteorological data were collected} \\ \cline{2-7}
& \multicolumn{3}{c}{Receiver location only} & \multicolumn{3}{|c}{10 locations near the propagation path} \\ \cline{2-7}
& \multicolumn{6}{c}{Number of meteorological factors applied} \\ \cline{2-7}
& 3 & 5 & \textbf{7} & 3 & 5 & \textbf{7} \\ \cline{2-7}
\hline
BPNN [ns] & \multicolumn{1}{|c}{50} & 46 & 38 & \multicolumn{1}{|c}{48} & 42 & 35 \\
GRNN [ns] & \multicolumn{1}{|c}{45} & 44 & 36 & \multicolumn{1}{|c}{38} & 37 & 34 \\
MoE [ns] & \multicolumn{1}{|c}{38} & 34 & 26 & \multicolumn{1}{|c}{32} & 29 & 23 \\
\textbf{LASSO-regularized MPR [ns]} & \multicolumn{1}{|c}{40} & 35 & \textbf{29} & \multicolumn{1}{|c}{32} & 31 & \textbf{26} \\
\textbf{WLR--AGRNN [ns]} & \multicolumn{1}{|c}{30} & 28 & \textbf{24} & \multicolumn{1}{|c}{22} & 20 & \textbf{17} \\
\thickhline
\end{tabular}
\end{table*}

\begin{table}
\centering
\caption{One-way ANOVA test results comparing RMSE and MAE across the five eLoran/GPS TD estimation models.}
\label{table:ANOVA}
\renewcommand{\arraystretch}{1.5}
\begin{tabular}{{w{c}{1.7cm} !{\vline} w{c}{1.7cm} w{c}{1.7cm} }}
\hline
Metric & RMSE & MAE \\
\hline
$p$ & <0.001 & <0.001 \\
\hline
\end{tabular}
\end{table}

We collected eLoran, GPS, and meteorological data over a four-month period, as described in Section~\ref{sec:DataCollection}. 
This significantly exceeds the validation periods of approximately one week reported in previous studies \cite{Pu2019:Analysis, Liu2023:eLoran, Pu2021:Accuracy}. 
Data from October, November, and late January were used for training, while data from December through early January were reserved for performance evaluation. 
A total of 198 points along the signal propagation path from the eLoran transmitter to the receiver were selected. 
Terrain elevation and meteorological data for these points were prepared using the methods outlined in Section~\ref{sec:Interpolation}.

For comparison, we constructed two existing estimation models---BPNN and GRNN---which have been used in previous studies to predict Loran-C propagation delay \cite{Pu2019:Analysis, Pu2021:Accuracy, Liu2023:eLoran}. 
Additionally, we implemented the basic MoE model \cite{Jacobs1991:Adaptive}, which, although not previously applied to eLoran/GPS TD estimation, served as the conceptual foundation for partitioning and learning input data types in the WLR--AGRNN model. 
The proposed models included in the comparison are the LASSO-regularized MPR and the WLR--AGRNN models.
During the training process of the WLR--AGRNN model, we employed the Adam optimizer with a learning rate of 0.001 and used full-batch training.

The computational complexity of both the BPNN and the basic MoE model can be expressed as $O(N \times H)$, where $H$ denotes the number of hidden neurons \cite{Lister1995:Empirical, Bienstock2023:Principled}. 
Since $H$ can be treated as constant in our experiments, the resulting complexity is $O(N)$. 
The computational complexity of the GRNN is also $O(N)$, consistent with the complexity derived for AGRNN in Section~\ref{sec:WLR--AGRNN2}.

Each model was trained using three different combinations of meteorological factors:
\begin{itemize}
    \item Three factors---temperature, humidity, and atmospheric pressure---as used in \cite{Pu2021:Accuracy}.
    \item Five factors---temperature, humidity, vapor pressure, visibility, and wind velocity---as used in \cite{Pu2019:Analysis}.
    \item Seven out of eleven selected factors---atmospheric pressure, cloud cover, humidity, temperature, vapor pressure, visibility, and wind velocity---as proposed in Section~\ref{sec:analysis}.
\end{itemize}

Previous studies used meteorological data collected either only at the receiver's location for training (${x^{(i)}_l}$), as in \cite{Pu2019:Analysis}, or at multiple locations along the signal propagation path (${x^{(i)}_1, x^{(i)}_2, \dots, x^{(i)}_l}$), as in \cite{Pu2021:Accuracy, Liu2023:eLoran}. 
To account for this, we evaluated the performance of each model under six different conditions, covering three combinations of meteorological factors and two variations in data collection locations.

We utilized two commonly used evaluation metrics---root mean square error (RMSE) and mean absolute error (MAE)---to assess the estimation accuracy of our eLoran/GPS TD models, as well as the baseline models. 
The mathematical definitions of these metrics are provided below:
\begin{equation}
    \text{RMSE} = \sqrt{\frac{1}{N}\sum_{i=1}^{N}(y_i - \hat{y}_i)^2}
\end{equation}
\begin{equation}
    \text{MAE} = \frac{1}{N}\sum_{i=1}^{N}|y_i - \hat{y}_i|
\end{equation}

RMSE and MAE are widely adopted for evaluating prediction accuracy in regression tasks. 
RMSE is more sensitive to large errors due to the squaring of residuals, making it particularly useful when identifying and penalizing substantial deviations is important. 
In contrast, MAE provides a straightforward measure of the average absolute error and is less affected by extreme outliers, offering a more robust representation of typical model performance. 
Together, these two metrics offer a comprehensive evaluation of the predictive capability and reliability of regression models.

\begin{figure}
  \centering
  \includegraphics[width=1.0\linewidth]{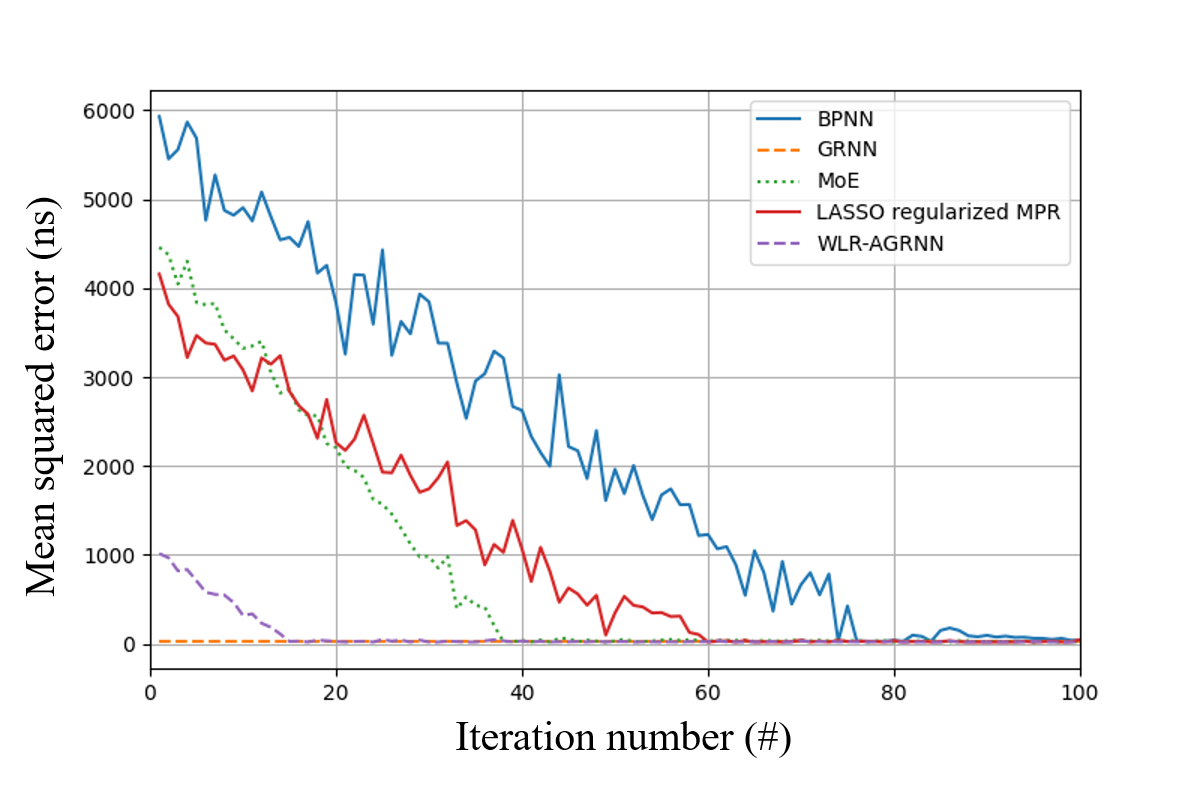}
  \caption{Convergence behavior of five eLoran/GPS TD estimation models in terms of estimation error across training iterations.}
  \label{fig:ConvGraph}
\end{figure}

Figure~\ref{fig:ConvGraph} shows the convergence behavior of five different eLoran/GPS TD estimation models. 
As illustrated in the figure, BPNN begins with the highest initial estimation error and requires the most iterations to converge. 
While MoE and the LASSO-regularized MPR model exhibit similar convergence trends, models such as GRNN and AGRNN---which do not rely on iterative training---require significantly fewer iterations to converge. 
In particular, GRNN does not require any iterative training, and therefore maintains the same estimation error from the first to the final iteration.
While this study empirically demonstrates the convergence behavior of the proposed model, deriving formal convergence guarantees for the integrated WLR--AGRNN framework remains an important avenue for future research.

Table~\ref{table:RMSE} summarizes the RMSEs, and Table~\ref{table:MAE} summarizes the MAEs for the estimated eLoran/GPS TD values across five different models, compared to the measured TD.
Both RMSE and MAE are evaluated under three different combinations of meteorological factors and two variations in meteorological data collection locations.
First, the results show that using seven meteorological factors yields higher estimation accuracy than using fewer factors, consistently across all five models. 
Second, TD estimation based on meteorological data collected at multiple locations along the signal propagation path is more accurate than estimation based solely on data from the receiver’s location, again consistently across all models. 
These observations align with expectations and are intuitively reasonable.

When trained with the same number of meteorological factors and data collection locations, BPNN produced the least accurate estimations. 
This can be attributed to several factors, including BPNN’s susceptibility to local minima, the absence of interaction terms, and a network structure that limits the effective integration of meteorological and terrain elevation data.

The LASSO-regularized MPR model outperformed GRNN by effectively capturing intercorrelations among input variables through interaction terms, particularly when all seven meteorological factors were included in the model training. 
In polynomial regression models, the number of input variables---represented by polynomial terms---increases with the inclusion of additional meteorological factors. 
Such models are known to perform well when the number of input variables is large \cite{Zhou2021:House}.

The MoE model demonstrated slightly better performance than the LASSO-regularized MPR model and significantly outperformed both BPNN and GRNN. 
This suggests that a composite network composed of multiple specialized subnetworks---each optimized for different meteorological factors---can yield more accurate TD estimations than a single, unified network.

Among the five models, the WLR--AGRNN model achieved the highest estimation accuracy for eLoran/GPS TD. 
Its superior performance can be attributed to its network architecture, which effectively integrates both meteorological and terrain elevation data---unlike the other models, which are designed to use only meteorological inputs.

Table~\ref{table:ANOVA} presents the statistical significance test results conducted using the one-way Analysis of Variance (ANOVA) method.  
ANOVA evaluates whether the performance differences among multiple models are statistically significant by comparing their variances and computing the resulting $p$-values \cite{Yu2022:Beyond}.  
For both RMSE and MAE, the $p$-values were found to be less than 0.05, indicating that the differences in performance among the estimation models are statistically significant.

\begin{figure}
  \centering
  \includegraphics[width=1.0\linewidth]{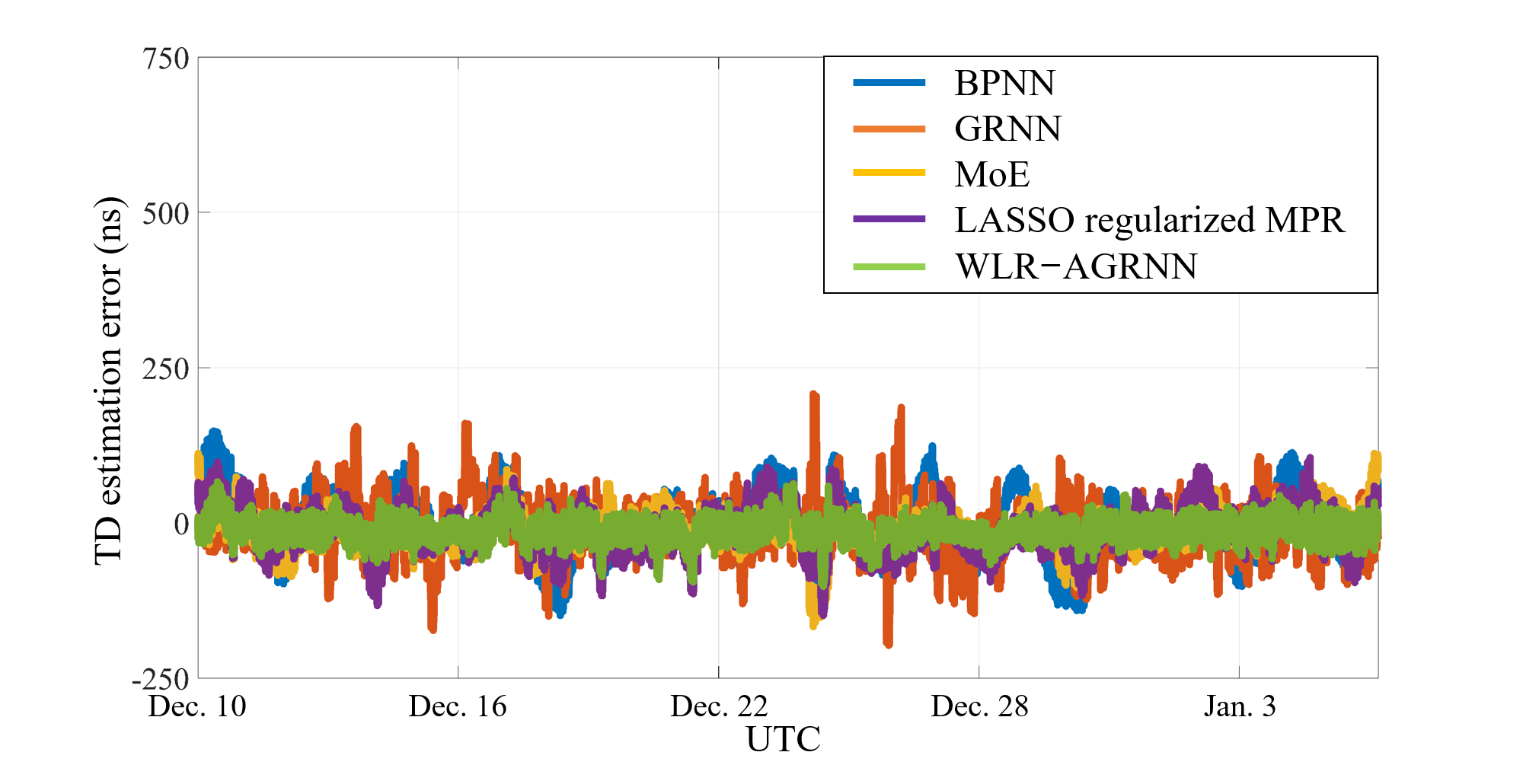}
  \caption{Comparison of eLoran/GPS TD estimation errors for five models over a one-month period. The WLR--AGRNN model shows the lowest and most consistent errors.}
  \label{fig:CorrCompare}
\end{figure}

Figure~\ref{fig:CorrCompare} compares the TD estimation errors of the five models over a one-month period. 
All models were trained using seven meteorological factors, with data collected from ten locations along the signal propagation path between the transmitter and the receiver. 
The figure clearly demonstrates that the WLR--AGRNN model, represented by the green line, consistently outperforms the other four models in terms of estimation accuracy.

\section{Conclusion}
\label{sec:Conclusion}

In this paper, we analyzed the correlations between eLoran/GPS TD and various meteorological factors and selected those with strong correlations to develop eLoran/GPS TD estimation models. 
To improve data accuracy at locations without direct measurements, we constructed grid maps of meteorological data using measurements from a limited number of stations along the groundwave propagation path. 
We proposed two learning-based estimation models---LASSO-regularized MPR and WLR--AGRNN---that achieve higher estimation accuracy than BPNN and GRNN. 
Notably, the WLR--AGRNN model is specifically designed to account for the influence of terrain elevation along the propagation path. 
Experimental results based on four months of data demonstrated that the WLR--AGRNN model consistently outperformed the other models, highlighting its effectiveness in improving eLoran/GPS TD estimation accuracy.  
By improving the accuracy of eLoran timing information, the proposed models can enhance the reliability of timing-dependent systems such as financial networks, transportation infrastructure, and safety-critical communication systems, particularly in GPS-denied environments.  
In addition to broader geographic validation, future research could explore extending the WLR--AGRNN framework for dynamic ASF estimation, offering an alternative to static ASF maps by leveraging real-time meteorological and topographic data.

\bibliographystyle{IEEEtran}
\bibliography{mybibliography, IUS_publications}

\phantomsection

\begin{IEEEbiography}[{\includegraphics[width=1in,height=1.25in,clip,keepaspectratio]{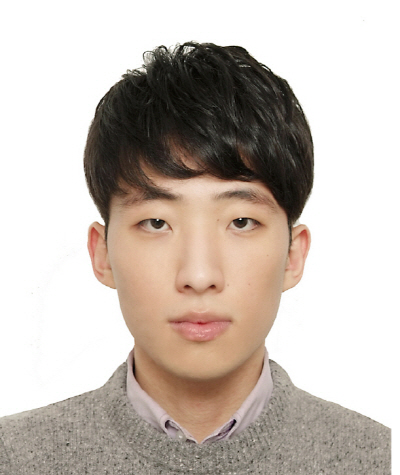}}]{Taewon Kang} received the B.S. degree in integrated technology from Yonsei University, Incheon, South Korea, in 2016, where he is currently pursuing the Ph.D. degree in integrated technology. 
He was a recipient of the Undergraduate and Graduate Fellowships from the ICT Consilience Creative Program supported by the Ministry of Science and ICT, South Korea. His research interests include complementary navigation systems, indoor navigation, and intelligent unmanned systems. 
\end{IEEEbiography}

\begin{IEEEbiography}[{\includegraphics[width=1in,height=1.25in,clip,keepaspectratio]{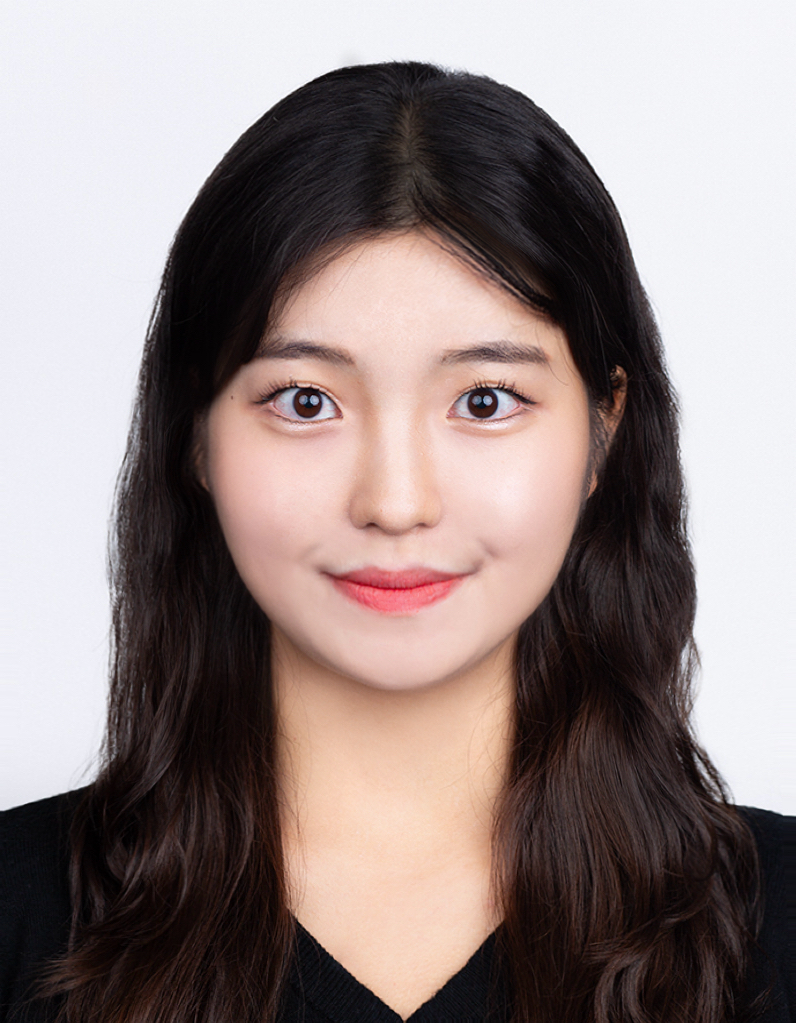}}]{Seunghyeon Park} received the B.S. degree in integrated technology from Yonsei University, Incheon, South Korea, in 2023, where she is currently pursuing the Ph.D. degree in integrated technology.
She was a recipient of the Talent Award of Korea in 2023, awarded by the Deputy Prime Minister and Minister of Education of the Republic of Korea.
Her research interests include complementary PNT systems, indoor navigation, and LEO satellite applications.
\end{IEEEbiography}

\begin{IEEEbiography}[{\includegraphics[width=1in,height=1.25in,clip,keepaspectratio]{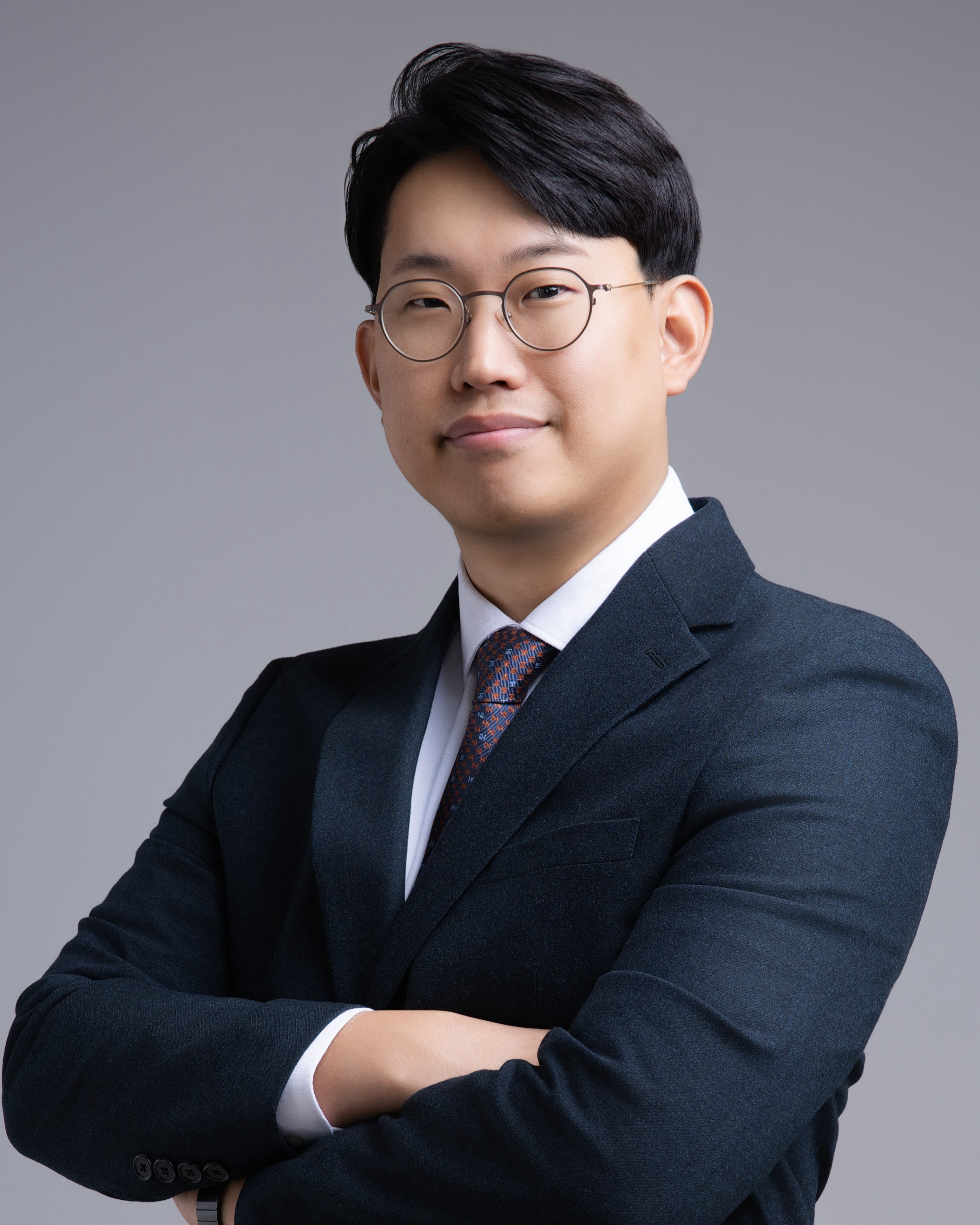}}]{Pyo-Woong Son} received his B.S. degree in electrical and electronic engineering from Yonsei University, Seoul, South Korea, in 2012, and his Ph.D. in integrated technology from Yonsei University, Incheon, South Korea, in 2019.
He is currently an Assistant Professor at Chungbuk National University, Cheongju, South Korea.
His research interests include complementary positioning, navigation, and timing systems, including eLoran.
\end{IEEEbiography}

\begin{IEEEbiography}[{\includegraphics[width=1in,height=1.25in,clip,keepaspectratio]{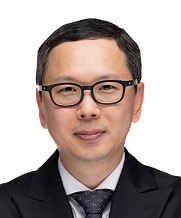}}]{Jiwon Seo} (Member, IEEE) received the B.S. degree in mechanical engineering (division of aerospace engineering) from the Korea Advanced Institute of Science and Technology (KAIST), Daejeon, South Korea, in 2002. 
He received M.S. degrees in aeronautics and astronautics (2004) and electrical engineering (2008), and the Ph.D. degree in aeronautics and astronautics (2010), all from Stanford University, Stanford, CA, USA.
He is currently an Underwood Distinguished Professor at Yonsei University, where he is a Professor in the School of Integrated Technology, Incheon, South Korea. 
He also serves as an Adjunct Professor in the Department of Convergence IT Engineering at Pohang University of Science and Technology (POSTECH), Pohang, South Korea.
His research interests include GNSS anti-jamming technologies, complementary positioning, navigation, and timing (PNT) systems, and intelligent unmanned systems. 
Dr. Seo is a member of the International Advisory Council of the Resilient Navigation and Timing Foundation, Alexandria, VA, USA, and serves on the Advisory Committee on Defense of the Presidential Advisory Council on Science and Technology, South Korea.
\end{IEEEbiography}

\EOD

\vfill

\end{document}